\documentclass[fleqn,superscriptaddress,showpacs,showkeys,twocolumn]{revtex4}
\usepackage{graphicx}
\usepackage{bm}
\usepackage{ulem}
\usepackage{dcolumn}
\newcommand{\Det}{{\rm Det}}
\newcommand{\intzr}{\int d^4 z d\rho\, d(\rho)}

\newcommand{\ave}[1]{\langle #1\rangle}
\newcommand{\intp}{\int \frac{d^4p}{(2\pi) ^4}}
\newcommand{\intk}{\int \frac{d^4k}{(2\pi) ^4}}
\newcommand{\intq}{\int \frac{d^4q}{(2\pi) ^4}}
\newcommand{\half}{{\frac{1}{2}}}
\newcommand{\Tr}{{\rm Tr}}

\newcommand{\bea}{\begin{eqnarray}}
\newcommand{\eea}{\end{eqnarray}}
\newcommand{\ba}{\begin{array}{l}}
\newcommand{\ea}{\end{array}}
\newcommand{\gf}{\gamma_5}

\newcommand{\re}[1]{(\ref{#1})}
\usepackage[usenames,dvipsnames]{color}

\begin{document}
\preprint{}
\title {Low energy constants of $\chi$PT from the instanton vacuum model}
\author{K. Goeke}
\email{Klaus.Goeke@tp2.rub.de}
\affiliation{Institut f\"ur  Theoretische Physik II, Ruhr-Universit\"at-Bochum, D-44780 Bochum, Germany}
\author{M.M. Musakhanov}
\email{yousuf@uzsci.net}
\affiliation{Theoretical  Physics Dept,Uzbekistan National University, Tashkent 700174, Uzbekistan}
\author{M. Siddikov}
\email{Marat.Siddikov@tp2.rub.de}
\affiliation{Theoretical  Physics Dept,Uzbekistan National University, Tashkent 700174, Uzbekistan}
\affiliation{Institut f\"ur  Theoretische Physik II, Ruhr-Universit\"at-Bochum, D-44780 Bochum, Germany}
\date{\today}

\begin{abstract}
In the framework of the instanton vacuum model we make expansion over the current mass $m$ and number of colors $N_c$ and evaluate ${\cal O}(1/N_c,\,m,\,m/N_c,\,m \,\ln m/N_c)$-corrections to the  dynamical quark mass $M$, the quark condensate $\langle\bar qq\rangle$, the pion mass $M_\pi$ 
and decay constant $F_\pi$. There are several sources of these corrections: meson loops, finite size of the instanton distribution and the quark-quark "tensor" interaction terms. In contrast to the expectations, we found that numerically the $1/N_c$-corrections to dynamical mass are large and mostly come from meson loops. As a consequence, we have large $1/N_c$-corrections to all the other quantities. To provide the values of $F_\pi(m=0), \ave{\bar qq(m=0)}$ in agreement with $\chi$PT, we offer a new set of parameters $\rho,\, R$. Finally, we find the low-energy $SU(2)_f$ chiral lagrangian constants $\bar l_3, \bar l_4$ in a rather good correspondence with the phenomenology.
\end{abstract}
\maketitle
 
 \section{Introduction}
 
 The spontaneous breaking of chiral symmetry (S$\chi$SB) is one of the most important phenomena  of hadron physics. It defines the properties of all the light mesons and baryons. Using the general idea of chiral symmetry, the Chiral Perturbation Theory ($\chi$PT) was proposed in~\cite{Gasser:1983yg}  for evaluation of the QCD hadronic correlators at low-energy region, where the expansion parameters are  light quark current masses $m$ and pion momenta $p$. The basic tool is the phenomenological effective lagrangian, which has a form of the infinite series in these parameters $p$ and $m$. Naturally, the low-energy constants (LEC) of the series expansion are not fixed. Up to now they were extracted only from the experimental data. Recent progress in lattice calculations provide {us with rough estimates of LEC. The main problem of lattice evaluations are the still-large pion masses $M_\pi$ available on the finite size lattices.}
 
 QCD instanton vacuum model, often referred  as the instanton liquid model, provides a very natural nonperturbative explanation of the S$\chi$SB\cite{Schafer:1996wv,Diakonov:2002fq}. It provides a consistent framework for description of the pions and thus may be used for evaluation of the LEC.

  Quasiclassical considerations show that it is energetically favourable to have lumps of strong gluon fields (instantons) spread  over 4-dimensional Euclidian space. Such fields do strongly modify the quark propagation due to the t'Hooft type quark-quark interactions in the background of the instanton vacuum field. This background is assumed as a superposition of $N_+$ instantons and $N_-$ antiinstantons 
 \begin{eqnarray}
 \label{A}
 A_\mu (x) = \sum^{N_+}_I A^I_{\mu}(\xi_I, x)
 +\sum^{N_-}_{A} A^{A}_{\mu}(\xi_{A}, x),
 \end{eqnarray}
 where $\xi=(\rho,z,U)$ are the (anti)instanton collective coordinates -- size, position and color orientation. The most essential for the low-energy processes are the would-be quark zero modes, which result in a very strong attraction in the channels with quantum numbers of vacuum, appearance of the quark condensate
 and generation of the dynamical quark mass (see the reviews ~\cite{Schafer:1996wv,Diakonov:2002fq}).  
 
 The main parameters of the model are the average inter-instanton distance $R$ 
 and the average instanton size $\rho$. The  estimates of these quantities are
\begin{eqnarray}
\label{rhoR}
 && \rho\simeq   0.33 \, fm, \, R\simeq 1\, {\rm fm},
\mbox{(phenomenological)}~ ~\mbox{\cite{Shuryak:1981ff}}, \label{classicalParameters}
\\
 && \rho\simeq 0.35\, fm,\, R \simeq 0.95\,{\rm fm}, 
 \mbox{(variational)}~ ~\mbox{\cite{Diakonov:1983hh}},
\nonumber\\
 &&\rho\simeq 0.36\,fm,\,   R\simeq 0.89\,{\rm fm},
~\mbox{(lattice)} ~\mbox{\cite{Chu:vi,Negele:1998ev,DeGrand:2001tm,Faccioli:2003qz,Bowman:2004xi}}
\nonumber
 \end{eqnarray}
 and have $\sim 10-15\%$ uncertainty.

Recent computer investigations~\cite{Negele0605256} of a current mass dependence of QCD observables within instanton liquid model show that the best correspondence with lattice QCD data is obtained for 
\begin{eqnarray}
\label{fromNegele0605256}
\rho\simeq  0.32\, fm,\,\,\,\, R\simeq  0.76\, fm.
\end{eqnarray}
\\
While in the real world the number of colors is $N_c=3$, since the pioneering work of t'Hooft~\cite{'tHooft:1973jz} it is \textit{assumed} that one can consider  $N_c$-counting  as a useful tool, \textit{i.e.} take the limit $N_c\to\infty$ and neglect the $1/N_c$-corrections. In the instanton vacuum model $N_c$-counting is naturally incorporated.
The phenomenological set~\re{classicalParameters} is popular since in the leading order (LO) on $N_c$ it reproduces reasonable values for most of the physical quantities. This leads to rather consistent description of pions and nucleons in the chiral limit.

The main purpose of this paper is evaluation of ${\cal O}(1/N_c,\,m,\,m/N_c,\,m/N_c\,\ln m)$-corrections to different physical observables.
There are several sources of such NLO corrections:\\
\begin{enumerate}
\item At pure gluonic sector of the instanton vacuum model the width of the instanton size distribution is ${\cal O}(1/N_c)$. The account of the finite width leads to rather small corrections~\cite{Diakonov:1985eg}. In the following we will check the accuracy of $\delta$-function type of the instanton size distribution by direct evaluation of the finite width corrections.
\\
\item The back-reaction of the light quark determinants to the instanton vacuum properties is formally controlled by $N_f/N_c$-factor~\cite{Diakonov:1995qy}. It does not sizably change the distribution over $N_++N_-$ but radically change the distribution over $N_+ - N_-$. Any $m_f=0$ leads to $\delta$-function type of the distribution ~\cite{Diakonov:1995qy}. In the following we take $N_+ = N_-$.
\\
\item {There are the quark-quark tensor interaction terms which are $1/N_c$-suppressed and thus are absent in the LO effective action. These terms correspond to nonplanar diagrams in old-fashioned diagrammatic technique.}
\\
\item The contribution of meson quantum fluctuations (meson loops) has to be taken into account.
\end{enumerate}
 In the first few sections we study the role of the meson loops which give the dominant contribution. At the end we also estimate the contributions of finite width of instanton size distribution, and  above-mentioned tensor interaction term.
\\
We consider parameters $\rho, R$ as free within their $\sim 15\%$ uncertainty and fix them from the requirement $F_{\pi,{m=0}}=88 MeV, \langle\bar qq\rangle_{m=0}=(255 MeV)^3$ with account of NLO corrections, as it is requested by $\chi$PT~\cite{Gasser:1983yg}. We found the values
 \begin{eqnarray}
 \rho=0.350 fm,\,\,R=0.856 fm,\label{Param:fit}
 \end{eqnarray}
 in agreement with the above-given estimates~(\ref{rhoR},\ref{fromNegele0605256}).

Note that though the evaluation of the meson loop corrections in the instanton vacuum model is similar to the earlier meson loop evaluations \cite{Nikolov:1996jj,Plant:2000ty,Pena:1999sp,Oertel:2000cw} in the NJL model, there are a few differences which should be mentioned:
\begin{enumerate}
\item As it has been already  mentioned, the meson loop corrections are not the only sources of $1/N_c$-corrections in the instanton model.

\item  Due to nonlocal formfactors there is no need to introduce independent fermion and boson cutoffs $\Lambda_f, \Lambda_b$. The natural cutoff scale for all the loops (including meson loops) is the inverse instanton size $\rho^{-1}$.

\item The quark coupling constant is defined through the saddle-point equation in the instanton model whereas it is a fixed external parameter in NJL.

\end{enumerate}

 \par In Section \ref{SectDerivation}, using the above mentioned assumptions and zero-mode approximation (see below),  we calculate the quark determinant in the gluon background $\re{A}$ and flavour vector $v_\mu $, axial-vector $ a_\mu$, scalar $s$ and pseudoscalar $p$  external fields beyond chiral limit. Since the zero-mode approximation leads to a violation of the flavour local gauge invariance, we formally restore it by introducing the path-ordered exponents. The price of such restoration is the path-dependence. We show, however, that the quantities considered in this paper do not depend on the choice of the path (see Sec. \ref{SectPathIndep}).

After that, we average  the quark determinant over instanton collective coordinates as it was described in~\cite{Musakhanov:1996qf,Salvo:1997nf,Musakhanov:1998wp}, using smallness of packing parameter $(\frac{\rho}{R})^4\sim 0.01.$ This leads to the partition function $Z_N [v,a,s,p,m],$ which is a generating functional of various quark correlators.
The partition function is represented as a path integral over the fermions, which are in the present formalism constituent quarks with t'Hooft-like $2N_f$-legs nonlocal interaction term  with the coupling constant $\lambda$, defined dynamically from the appropriate  equation. In the following we consider only $N_f=2$ case.

The simplest and the most important quark correlator is the quark propagator, since all the other observables  are sensitive to its properties. In section~\ref{SectDynMass} we calculate the dynamical quark mass $M$.
We found in the instanton vacuum model that contrary to the large-$N_c$ expectations, the meson-loop corrections to $M(m)$ are large ($\sim 50\%$) even in the chiral limit. These corrections are intimately related to the equation for the coupling $\lambda$. Such corrections are absent in NJL-type models.

All meson-loop corrections to physical observables can be splitted into two parts: "indirect" (which are dominant and come from the dynamical mass shift) and "direct" (which are relatively small and thus obey $1/N_c$-counting rules.)

 In sections \ref{SectQuarkCondensate}-\ref{SectAAcorrelator} we calculate the meson-loop corrections to the quark condensate $\langle\bar qq\rangle$, pion mass $M_\pi$ and decay constant $F_\pi$. From ${\cal O}(m,1/N_c)$ corrections to the latter quantities in section \ref{SectGassCoupl} we extract the low-energy parameters $\bar l_3,\bar l_4$ of the phenomenological Gasser-Leutwyler $SU(2)_f$ lagrangian~\cite{Gasser:1983yg}. These two constants are especially important for chiral perturbation theory \cite{Leutwyler:2006qq,Leutwyler:2007ae} since they relate the physical observables $F_\pi, M_\pi$ with their parameters. In section~\ref{SectChiralLogs} we check that the chiral log theorems~\cite{LangackerPagels73,Novikov:xj,Gasser:1983yg}  are satisfied.
   In section~\ref{SectPathIndep} we demonstrate that our results are path independent. Appendix \ref{SectVertices} contains explicit expressions of the nonlocal vertices of vector and axial currents. In Appendix \ref{SectFWCDetails} we give details of evaluation of finite width corrections, and in Appendix \ref{Sect:Transversity} we study the structure of the correlator $\ave{j_\mu^{a,5}j_\nu^{b,5}}$.

 \section{Light quarks  partition function in external fields}
 \label{SectDerivation}
 The light quark  partition function $Z[m,v,a,s,p]$ with quark mass $m$ and in external vector $v_\mu $, axial-vector $ a_\mu$, scalar $s$ and pseudoscalar $p$ fields is defined as
 \begin{eqnarray}
 \label{GenFunc}
 &&Z[v,a,s,p,m]
\\\nonumber
&&=\int D\,A_\mu\,e^{-\frac{1}{4}F^2}\det(i\hat D+im+\hat v+\hat a \gf+s+p\gf).
 \end{eqnarray}
 The basic assumption of the instanton model is that one can evaluate the integral in quasiclassical approximation, making expansion around the classical vacuum   \re{A}. The first evaluation of the partition function \re{GenFunc} was performed in Ref.~\cite{Diakonov:1985eg,Diakonov:1995qy} in the absence of the external fields and in the chiral limit.
 The main purpose of the present paper is the extension of the result of \cite{Diakonov:1985eg,Diakonov:1995qy} to the case of nonzero quark mass $m$ and external $v,a,s,p$ fields. While inclusion of the spin-$0$ fields $s,\vec p$ is trivial, spin-$1$ fields $v_\mu,a_\mu$ deserve special attention due to the nonhomogeneous transformation under flavour rotations. 
 Following ~\cite{Diakonov:1985eg,Diakonov:1995qy} we split the quark determinant into the low- and high-frequency parts according to $\Det =\Det_{\rm high}\cdot \Det_{\rm low}$ and  concentrate on the evaluation of $\Det_{\rm low}$, which is responsible for the low-energy domain. The high-energy part  $\Det_{\rm high}$ is responsible mainly for the perturbative coupling renormalization.
 \par The starting point of our consideration is the zero-mode approximation formulated in ~\cite{Lee:sm,Diakonov:1985eg,Diakonov:1995qy}
 \begin{eqnarray}
 \label{SiDP}
  S_i=\frac{1}{\hat p + \hat A_i +im}=\frac{1}{\hat p} 
  + \frac{|\Phi_{i,0}\rangle\langle\Phi_{i,0}|}{im}.
 \end{eqnarray}
Here the zero-modes $\Phi_{i,0}$ are also functions of the instanton collective coordinates $\xi_i.$
 This approximation is good for small values of $m$ (chiral limit) and indicates that the main contribution to the quark propagator is due to the zero-modes. The extension of~\re{SiDP} beyond the chiral limit was proposed in our previous 
 works ~\cite{Musakhanov:1998wp,Musakhanov:vu,Musakhanov:2002xa,Kim:2004hd} as
 \begin{eqnarray}
 \label{Si}
 S_i=S_{0} + S_{0}\hat p \frac{|\Phi_{0i}\rangle\langle\Phi_{0i}|}{c_i} \hat p S_{0}
 \end{eqnarray}
 where
 \begin{eqnarray}
 c_i=-\langle\Phi_{0i}|\hat p S_{0} \hat p |\Phi_{0i}\rangle .
 \end{eqnarray}
 The advantage of this approximation is that  it gives correct projection of $S_i$ to the zero-modes:
 \begin{eqnarray}
 S_i|\Phi_{0i}\rangle = \frac{1}{im}|\Phi_{0i}\rangle,\,\,\, \langle\Phi_{0i}|S_i
 =\langle\Phi_{0i}|\frac{1}{im},
 \end{eqnarray}
 while the similar projection of $S_i$ given by Eq.~\re{SiDP} is valid only at  $m\rightarrow 0$ limit.

Recently the quark determinant $\Det(\hat p+\hat A_i +im)=\exp(\Gamma_f)$ was calculated in the field of a single instanton and beyond the chiral limit~\cite{Carlitz:1978yj,Lee:2005}. The $\Gamma_f$ can be represented as $\Gamma_f=-\ln m\rho-2\Gamma_s$, where we take the scale equal to $1/\rho$.
Here the first term is due-to the zero-mode. At small $m\rho$ the
$\Gamma_s=0.145873+0.5 (m\rho)^2\ln(m\rho)-0.0580(m\rho)^2$. First, independent on $m$ term was calculated at~\cite{'tHooft:1976fv}. 
It is especially interesting for us the $m$-dependent terms in $\Gamma_s$. In the region $m\rho\sim 0.2$ (corresponding to the strange current quark mass) the third term
is of order $0.2\%$ of the second one and is negligible. So, the main problem is to reproduce the second term in $\Gamma_s$. Simple calculations with using the Eq. \re{Si} lead to the $\Gamma_s=(0.5\ln m\rho + 0.0767)(m\rho)^2$. Again, at the region $m\rho<0.2$ we may neglect by $(m\rho)^2$ term. So, our approximation \re{Si} leads to  the main $(m\rho)^2\ln m\rho$-term in $\Gamma_s$, which coincide with  exact calculations~\cite{Carlitz:1978yj,Lee:2005} in a sharp contrast to the zero-mode approximation \re{SiDP}, which leads to $\Gamma_s=0.$

 We use the notations 
 \begin{eqnarray}
 \tilde S = \frac{1}{\hat{p} +  \hat{A}  +\hat{V} +i{m}},&&
 \tilde S_i=  \frac{1}{\hat{p} +  \hat{A_i}  +\hat{V} +i{m}},\\
 \tilde S_0=\frac{1}{\hat{p} + \hat{V} +i{m}},&&
 S_0=\frac{1}{\hat{p} +i{m}}
 \end{eqnarray}
 for the quark propagators in the instanton background \re{A} or individual instantons $\hat A_i$ and in the external flavour fields $\hat V=\hat v + \hat a \gamma_5 + s + p\gamma_5,$ assumed to be weak. Our purpose is to evaluate the quark propagator $\tilde S$, which is just the inverse of the argument of $\Det$ in \re{GenFunc}.
 \par We can expand the quark propagator $\tilde S$ with respect to a single
 instanton field $A_i$:
\begin{eqnarray}
\tilde S=\tilde S_0+\sum_i (\tilde S_i-\tilde S_0)+\sum_{i\not=j} (\tilde S_i-\tilde S_0)\tilde S^{-1}_0(\tilde S_j-\tilde S_0)+\ldots
\label{S-tot}
\end{eqnarray}
 Now we are going to make expansion over $V$ and express $\tilde S_i$ via $S_i$. Notice that the field $\hat V$ transforms nonhomogeneously. If we could evaluate $S_i$ exactly, the final result would be gauge-covariant. However, due to the zero-mode approximation expansion over $V$ would kill the gauge-covariance of the propagator $\tilde S_i$. In order to preserve it, we introduce auxiliary field \begin{eqnarray}\hat{V_i^\prime}=\bar L_i(\hat{p}+\hat{V})L_i-\hat p,\label{covariantCurrent}\end{eqnarray}
 and the gauge connection $L_i$, which is defined as a path-ordered exponent
 \begin{eqnarray}
 &&L_i(x,z_i)={\rm P} \exp\left(i\int_{z_i}^x dy_\mu( v_\mu(y)+a_\mu(y)\gamma_5)\right),
 \nonumber\\
 &&\bar L_i(x,z_i)=\gamma_4 L_i^\dagger(x,z_i)\gamma_4
 \label{transporter}
 \end{eqnarray}
 where $z_i$ denotes an instanton position.
 The field $V'_i(x,z_i)$ under flavour rotation $$\psi(x)\rightarrow U(x)\psi(x)$$ transforms  according to $$V'_i(x,z_i)\rightarrow \bar U^{-1}(z_i)V'_i(x,z_i)U^{-1}(z_i)$$
 and thus expansion over it does not violate the gauge covariance.
 The propagators $\tilde S_i$ and $\tilde S_0$ in terms of the field $V_i$ have a form:
 \begin{eqnarray}
 \tilde S_i&=&L_iS'_{i}\bar L_i,\;\; S'_{i}=\frac{1}{\hat{p}
 +\hat{A_i} +\hat{V_i'}+i{m}},\cr
 \tilde S_0&=&L_iS'_{0i}\bar L_i,\;\;
 S'_{0i}=\frac{1}{\hat{p} +  \hat{V_i'}+i{m}}, 
 \end{eqnarray}
 Expanding $S'_{i}$ over $\hat V_{i}'$ and re-summing it, we get
 \begin{eqnarray}
&&S'_{i} = S_i(1+\sum_n (-\hat V_{i}' S_i)^n)=\\
&&\nonumber S'_{0i} + S'_{0i}\hat p\frac{|\Phi_{0i}\rangle\langle\Phi_{0i}|}{c_i - b_i} \hat p S'_{0i}
 \label{Si'}
 \end{eqnarray}
 where 
 \begin{eqnarray}
 \label{bi}
 &&b_i = \langle\Phi_{0i}|\hat p (S'_{0i}-S_0) \hat p |\Phi_{0i}\rangle,
 \\\nonumber
 &&c_i - b_i =-\langle\Phi_{0i}| \hat p S'_{0i} \hat p|\Phi_{0i}\rangle= 
 \\\nonumber
 &&\langle\Phi_{0i}| (im + \hat V_{i}')|\Phi_{0i}\rangle
  -\langle\Phi_{0i}|(im + \hat V_{i}') S'_{0i} (im + \hat V_{i}')|\Phi_{0i}\rangle
 \end{eqnarray}
 Rearrangement of the Eq. \re{S-tot} for the total propagator yields
 \begin{widetext}
 \begin{eqnarray}
\tilde S &=&\tilde S_{0} + \tilde S_{0}\sum_{i,j}
 \bar L_i^{-1} \hat p |\Phi_{i0}\rangle\left(\frac{1}{-D}+\frac{1}{-D}C\frac{1}{-D}+\ldots\right)_{ij}
 \langle\Phi_{0j}|\hat p L_j^{-1}\tilde S_{0}=\nonumber\\
 &=&\tilde S_{0} + \tilde S_{0}\sum_{i,j}\bar L_i^{-1}\hat p |\Phi_{i0}\rangle
 \left(\frac{1}{-V-T}\right)_{ij}\langle\Phi_{0j}|\hat p L_j^{-1}\tilde S_{0}
 \label{propagator}
 \end{eqnarray}
 \end{widetext}
 where
 \begin{eqnarray}
 &&V_{ij}=\langle\Phi_{0i}|\hat p (L_i^{-1}\tilde S_{0}\bar L^{-1}_j) \hat p|\Phi_{0j}\rangle -\langle\Phi_{0i}|\hat p S_{0} L_i^{-1}L_j \hat p |\Phi_{0j}\rangle,\nonumber\\
&&T_{ij}=(1-\delta_{ij})\langle\Phi_{0i}|\hat p S_0 L_i^{-1}L_j\hat p |\Phi_{0j}\rangle,\nonumber\\
&&D_{ij}=\delta_{ij}V_{ij}\equiv (b_i-c_i)\delta_{ij},\nonumber\\
&&C_{ij}=(1-\delta_{ij})V_{ij}.
\end{eqnarray}
It is natural to introduce
\begin{eqnarray}
|\phi_0\rangle=\frac{1}{\hat p}L \hat p |\Phi_0\rangle
\end{eqnarray}
which has the same chiral properties as zero-mode function $|\Phi_0\rangle$.
Then
\begin{eqnarray}
\tilde S -\tilde S_{0} = -\tilde S_{0}\sum_{i,j}\hat p |\phi_{0i}\rangle
\langle\phi_{0i}|\left(\frac{1}{V+T}\right)|\phi_{0j}\rangle
\langle\phi_{0j}|\hat p \tilde S_{0}
\label{propagator1}
\end{eqnarray}
with
\begin{eqnarray}
V+T=\hat p \tilde S_{0}\hat p.
\end{eqnarray}
From \re{propagator1} we get
\begin{eqnarray}
\Tr (\tilde S -\tilde S_0 )=-\sum_{i,j}\langle\phi_{0,j}|\hat p{\tilde S_{0}}^2 
\hat p |\phi_{0,i}\rangle\langle\phi_{0,i}|\left(\frac{1}{\hat p\tilde S_{0}\hat p}\right)|\phi_{0,j}\rangle
\nonumber
\end{eqnarray}
Introducing now the operator
\begin{eqnarray}
&&\tilde B(m)_{ij}^{fg}=\langle\phi_{0,i}|(\hat p \tilde S_0  \hat p)|\phi_{0,j}\rangle^{fg}=\\
&&\nonumber\langle\Phi_{0i}|\hat p  \left(L_{i}^{-1} \,\tilde S_{0}\bar L_{j}^{-1}\right)^{fg}\hat p |\Phi_{0j}\rangle,
\end{eqnarray}
where indices $i,j$ refer to the number of instantons and $f,g$ are flavour indices, it is easy to show that 
\begin{eqnarray}
&&\ln{\left(\Det_{low}\right)}=i\Tr\int^mdm'(\tilde S(m')- \tilde S_{0}(m'))=\\
&&= \sum_{i,j} \int^m dm 
\frac{d \tilde B(m')_{ij}}{dm}(\tilde B(m'))^{-1}_{ji}
=\tilde\Tr \ln {\tilde B(m)}\nonumber
\end{eqnarray}
Thus we have
\begin{eqnarray}
 {\Det}_{\rm low} [v,a,s,p,m] \cong \det \tilde B(m)
\label{tildeB}
\end{eqnarray}
where $\tilde B$ is the extension of Lee-Bardeen's matrix $B$~\cite{Lee:sm} for the case of nonzero external sources $ (v,a,s,p)$ and current quark mass~$m$. Certainly, in the limit of small $m$ and zero sources $\tilde B$ coincides  with $B$.

Independent averaging of the $\Det_{\rm low}$ over instanton collective coordinates $\xi ,$ justified in the dilute gas approximation and performed with the help of fermionization~\cite{Musakhanov:1996qf,Salvo:1997nf,Musakhanov:1998wp} gives the desired light quarks partition function $Z_N[v,a,s,p,m]$:
 \begin{widetext}
 \begin{eqnarray}
 Z_N=\ave{{\Det}_{\rm low} [v,a,s,p,m]}=\ave{\det \tilde B }
 &=& \int \prod_{f}D\psi_f D\psi^{\dagger}_{f} \exp\left(\int d^4 x
 \sum_{f,g}\psi_{f}^{\dagger}(\hat p \,+\,\hat V\,+\, im )_{fg}\psi_{g}\right)
\prod_{\pm}^{N_{\pm}} W_{\pm,f}[\psi^{\dagger} ,\psi ]\; ,
 \label{tildeBform}
 \end{eqnarray}

 where
\begin{eqnarray}
W_\pm[\psi^{\dagger} ,\psi]=\int d\xi_\pm \prod_{f}V_{\pm,f}[\psi^{\dagger} ,\psi ]
\label{W}
\end{eqnarray}
are the t'Hooft-like non-local interaction term with $N_f$ pairs of quark legs,

 \begin{eqnarray}
 \tilde V_{\pm}[\psi^{\dagger} ,\psi]=
 \int d^4 x \left(\psi^{\dagger} (x)\,\bar L^{-1}(x,z_\pm)\, \hat p
 \Phi_{\pm , 0} (x; \xi_{\pm})\right)\int d^4 y\left(\Phi_{\pm , 0}^\dagger (y; \xi_{\pm} )
 (\hat p\, L^{-1}(y,z_\pm) \psi (y)\right),
 \label{tildeV}
 \end{eqnarray}
 \end{widetext}
 and gauge links are defined in \re{transporter}.

 The form-factor of the nonlocal interaction is completely defined by the quark zero-mode.
The fermion fields $\psi^{\dagger} ,\psi$ are interpreted as constituent quarks. 
For the exponentiation we use Stirling-like formula
\begin{eqnarray}
W_\pm^{N_\pm}=\int d\lambda_\pm \exp(N_\pm\ln\frac{N_\pm}{\lambda_\pm}-N_\pm+\lambda_\pm W_\pm),
\label{lambda}
\end{eqnarray}
where $\lambda_\pm$ play a role of the dynamical coupling constant, defined by saddle-point condition at the integral \re{lambda}. Notice that the validity of the formula~(\ref{lambda}) is  controlled by the large number of instantons $N_\pm$.

The partition function in the instanton vacuum  after integration over collective coordinates and exponentiation may be reduced to the form (for the case $N_f=2$):
\begin{eqnarray}
&&Z_N=\int d\lambda_+d\lambda_-D\bar\psi D\psi e^{-S}\label{Z:withTensorTerms}\\
&&S=N_\pm \ln\frac{K}{\lambda_\pm}-N_\pm+\psi^\dagger (i\hat \partial+\hat V+im)\psi + \lambda _\pm Y_2^\pm \\
&&Y_2^\pm=\alpha^2 \det_f J^\pm+\beta^2 \det_f J^\pm_{\mu\nu}\label{Y2:definition}\\
&&\frac{\beta^2}{\alpha^2}:= \frac{1}{8N_c}\frac{2 N_c}{2 N_c-1}=\frac{1}{8N_c-4}={\cal O}\left(\frac{1}{N_c}\right)\label{TensorCouplingSuppression}\\
&&J^\pm=\psi^{\dagger '} \frac{1\pm \gf}{2}\psi',\\
&&J^\pm_{\mu\nu}=\psi^{\dagger '} \sigma_{\mu\nu}\frac{1\pm \gf}{2}\psi'=\half\left(J_{\mu\nu}\pm\frac{i}{2}\epsilon_{\mu\nu\rho\lambda}J_{\rho\lambda}\right),\\
&&\;\mbox{  since } \sigma_{\mu\nu}=\frac{i}{2}\epsilon_{\mu\nu\rho\lambda}\sigma_{\rho\lambda}\label{gf}\\
&&J_{\mu\nu}=\psi^{\dagger '} \sigma_{\mu\nu}\psi'\label{J}\\
&& \psi^{\dagger'}=\psi^\dagger \bar L,\;\;\psi'=L^{-1}\psi\\
&& L=\exp\left(i\int V_\mu(\xi)d\xi_\mu\right),\;\; \bar L=\gamma_0 L^\dagger \gamma_0
\end{eqnarray}
where the determinant is taken over the (implicit) flavour indices, and $K$ is some inessential constant introduced to make the argument of logarithm dimensionless.
From (\ref{TensorCouplingSuppression}) one can clearly see that the contribution of the tensor terms is just a $1/N_c$-correction. For the sake of simplicity we will postpone consideration of the tensor terms contribution until the Section \ref{SectTensorTerm} and concentrate on the first term in~(\ref{Y2:definition}). 

Next step is the bosonization of the interaction term $W_\pm[\psi^{\dagger} ,\psi]$.
 Here we are considering only the $N_f=2$ case, for which bosonization is an exact procedure, and take $N_+=N_-$, as discussed before. So, we get the partition function
 \begin{widetext}
 \begin{eqnarray}
 &&Z_N [v,a,s,p,m]
\nonumber\\
&&=\int d\lambda D\Phi D\bar\psi D\psi 
 \exp\left[N\ln\frac{K}{\lambda}-N + \int dx\left(-2
 \Phi^2 +\psi^\dagger(\hat p
 +\hat V+im+i\frac{(2\pi\rho)^2\lambda^{0.5}}{2g}\bar L \hat F
 \Phi\cdot \Gamma \hat F L)\psi\right)\right]\nonumber\\
 &&\mbox{or, integrating over fermions,}\nonumber\\
 \label{Z2}
 &&=\int d\lambda D\Phi e^{-S[\lambda,\Phi,v,a,s,p,m]}
  \\
&&\mbox{where,}\nonumber\\
 \label{S}
 \label{V0}
&&-S[\lambda,\Phi,v,a,s,p,m]=N\ln\frac{K}{\lambda}-N- 2\int dx
\Phi^2 +\Tr\ln\left(\frac{\hat p
 +\hat V+im+i\frac{(2\pi\rho)^2\lambda^{0.5}}{2g}\bar L \hat F
 \Phi\cdot \Gamma \hat F L}{\hat p +\hat V+im}\right)
 \end{eqnarray}
 \end{widetext}
 where $g^2 = \frac{(N_{c}^{2}-1)2N_c}{2N_c -1}$ is a color factor. Here   $\Gamma=\{1,\gamma_5,i\vec\tau,i\vec\tau\gamma_5\}$ correspond to the decomposition $\Phi=\{\Phi_0,\vec\Phi\}=\{\sigma,\eta,\vec\sigma,\vec\phi\}$, and $\Phi^2=\Phi_0^2+\vec\Phi^2=\sigma^2+\eta^2+\vec\sigma^2+\vec\phi^2$.
 Notice that in \re{Z2} 
in contrast to the NJL model the coupling constant $\lambda$ is  not a parameter of the action but it is defined by saddle-point condition in the integral \re{lambda}.

 Note that the partition function \re{Z2} is invariant under local flavour rotations due to the gauge links $L$ in the interaction term $\tilde V_{\pm, f}[\psi^{\dagger} ,\psi ]$. However, instead of the explicit violation of the gauge symmetry due to the zero-mode approximation (\ref{SiDP},\ref{Si}), we have unphysical dependence of the effective action on the choice of the path in the gauge link $L$.
 In our evaluations we used the simplest straight-line path, though there is no physical reasons why the other choices should be excluded. In Sec.~\ref{SectPathIndep} we demonstrate explicitly that for the quantities evaluated in this article the path dependence drops out.

 \section{Dynamical quark mass}
 \label{SectDynMass}
 One of the main advantages of the instanton vacuum model is the natural description of the  S$\chi$SB, which is signalled by non-zero vacuum quark condensate $\langle\bar qq\rangle$. The quark-quark interaction term \re{W} leads to the strong attraction in  the channels with vacuum (and pion) quantum numbers. As a consequence, there appear the nonzero vacuum expectation $\sigma$ of scalar-isoscalar component of meson fields $\Phi$ and related with it $\langle\bar qq\rangle$.
 For evaluation of the partition function $Z[m]$, it is very convenient to use the formalism of the effective action \cite{Coleman:1973jx,Jackiw74} $\Gamma_{eff}[m,\lambda,\Phi]$, defined as:
\begin{eqnarray}
&&Z_N[m]=\int d\lambda Z_N[m,\lambda]=\nonumber\\
&&\int d\lambda\exp(-\Gamma_{eff}[m,\lambda,\Phi])
\label{Veff}
\end{eqnarray} 
where for the sake of simplicity we dropped all the external currents which are not essential in this section,
and the field $\Phi$ is the solution of the vacuum equation
\begin{eqnarray}
\frac{\partial \Gamma_{eff}[m,\lambda,\Phi]}{\partial\Phi}=0.
\label{vacuum}
\end{eqnarray}
Notice that the solution depends on $\lambda$, \textit{i.e.} $\Phi=\Phi(\lambda)$. Up to the Section~\ref{SectAAcorrelator} it will be assumed that the only nonzero vacuum field is a condensate $\Phi=\sigma$, which is independent of coordinates, so the effective action $\Gamma_{eff}[m,\lambda,\Phi]$ may be replaced with effective potential $V_{eff}[m,\lambda,\sigma]$.

In the leading order, the effective action just coincides with the action (\ref{S}). Shifting $\Phi\to \sigma+\Phi'$ and integrating over the fluctuations, we get for the meson loop correction
\begin{widetext}
\begin{eqnarray}
\label{Vmes}
\Gamma_{eff}^{mes}[m,\lambda,\sigma]=\frac{1}{2}\Tr\ln\left( 4\delta_{ij}-\frac{1}{\sigma^2}
\Tr \frac{M(p)}{\hat p+ i \mu(p)} \Gamma_i \frac{M(p)}{\hat p+ i \mu(p) }\Gamma_j\right),
\end{eqnarray}
\end{widetext}
where $\mu(p)=m+M(p)$ and we introduced the dynamical quark mass $M(p)=M F^2(p)$; $M=\frac{(2\pi\rho)^2\lambda^{0.5}}{2g}\sigma.$

 It is convenient to introduce notations for the leading order meson propagators
 \begin{eqnarray}
 \label{Meson_Propagators}
 \Pi_i^{-1}(q)= 4 + \frac{1}{\sigma^2} V_2^i(q)
 \end{eqnarray}
 where
 \begin{eqnarray}
 \label{V2def}
 V_2^i(q)=Tr\left(Q(p)\Gamma_iQ(p+q)\Gamma_i\right)
 \end{eqnarray}
 and
 \begin{eqnarray}
 \label{Q(p)}
 Q(p)=S(p)iM(p)\equiv \frac{iM(p)}{\hat p+i\mu(p)}.
 \end{eqnarray}
 With these notations vacuum equation \re{vacuum} turns into
 \begin{eqnarray}
 \label{gap2}
 \sigma\frac{\partial \Gamma_{eff}}{\partial\sigma}=
 4\sigma^2-\frac{1}{V}Tr\left(Q(p)\right)&-&\\
 \nonumber \frac{1}{\sigma^2} \intq &\sum_i&V_3^i(q)\Pi_i(q)=0 ,
 \end{eqnarray}
 where
 \begin{eqnarray}
 \label{V3def}
 V_3^i(q)=Tr\left(Q^2(p)\Gamma_iQ(p+q)\Gamma_i\right).
 \end{eqnarray}
In the leading order (LO) over $N_c$ this equation simplifies to
 \begin{eqnarray}
 \label{gap2leading}
 4\sigma^2=\frac{1}{V}Tr\left(Q(p)\right).
 \end{eqnarray}

The equations (\ref{gap2},\ref{gap2leading}) completely define $\sigma, \sigma_{LO}$ as functions of $\lambda$. Using (\ref{gap2leading}), we may cast (\ref{Meson_Propagators}) into the form   
\begin{eqnarray}
 \Pi_i^{-1}(q)\Rightarrow \tilde\Pi_i^{-1}(q)=\frac{1}{\sigma^2 V}\left(Tr Q(p)+  V_2^i(q)\right)
 \label{tilde_Meson_Propagators}
\end{eqnarray}
This replacement guarantees self-consistency of the 1-meson-loop approximation and complies with Goldstone theorem:
\begin{eqnarray}
\tilde\Pi_\phi^{-1}(0)&=& \frac{8\,m\, N_c}{\sigma^2}\intp\frac{M(p)}{p^2+\mu^2(p)}
\\\nonumber
&=&\frac{2m\langle\bar qq\rangle (m=0)|_{LO} }{\sigma^2}+O(m^2).
\end{eqnarray}
Notice that the pion propagator has a correct chiral pole if and only if the field $\sigma$ satisfies the LO  equation (\ref{gap2leading}). Similar property is observed in NJL model (see, e.g., \cite{Nikolov:1996jj} and references therein).

There is an important difference between the instanton vacuum and
traditional NJL-type models -- the coupling $\lambda$ is not an external parameter of the model, but is defined from the saddle-point equation. 
We have to integrate over the coupling $\lambda$ in Eq. \re{Veff} to obtain partition function $Z_N$. The saddle-point approximation for the result becomes exact in the large-$N$ limit (see discussion of (\ref{lambda}) above). The saddle-point equation for $\lambda$ has a form
 \begin{eqnarray}
 \label{gap1}
  \label{gap_equations}
 \lambda\frac{\partial \Gamma_{eff}}{\partial\lambda}&=&
 \frac{N}{V}-\frac{1}{2V} Tr\left(Q(p)\right)+
 \\
 \nonumber
 &+&\frac{1}{2\sigma^2}\intq \sum_i\left(V_2^i(q)-V_3^i(q)\right)\Pi_i(q)=0
 \end{eqnarray}
 Notice that  $V_2$-term in \re{gap1} requires special attention. Formally it is next to leading order correction, while numerically it is strongly enhanced (about a factor of $30$ compared to the other $1/N_c$-corrections), which indicates the failure of the large-$N_c$ expansion in~(\ref{gap_equations}).

 \par If we solve the equation \re{gap1}, expanding it in powers of $1/N_c$ with the set of parameters \re{Param:fit},  we'll get
 \begin{eqnarray}
 \label{gap1_solution}
 &&M_0=0.567-2.362\,m\\
 &&M_1=\frac{1}{N_c}(-0.687 - 0.808\, m - 4.197 \, m\ln m ) .
\nonumber
 \end{eqnarray}
Here and in the following $M$ and $m$ are given in $GeV.$

 We can see that the meson loop correction $M_1$ is of comparable size with the LO term $M_0$, so we can try to \textit{solve the equations (\ref{gap2},\ref{gap1}) numerically in chiral limit and then evaluate the chiral corrections to it}. Such procedure gives
 \begin{eqnarray}
 \label{gap1_solution_exact}
 M(m)=0.36-2.36\,m -\frac{m}{N_c}(0.808+4.197 \ln m)
 \end{eqnarray}
The accuracy of the solutions(\ref{gap1_solution},\ref{gap1_solution_exact}) 
is ${\cal O}(m^2,\frac{1}{N_c^2}).$
 \begin{figure}[t]
\includegraphics[scale=0.4]{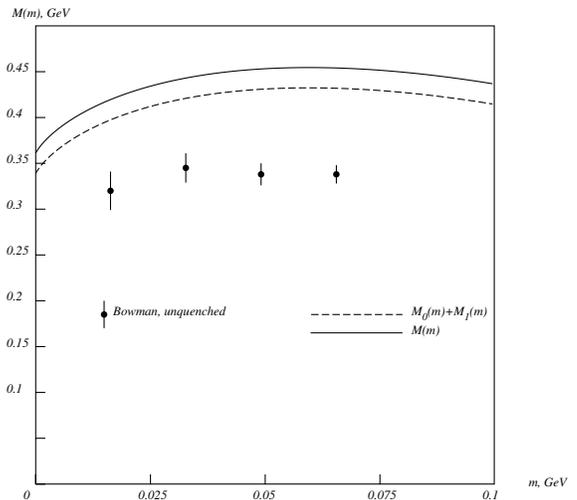}
 \caption{$m$-dependence of the dynamical quark mass $M$.
 The solid curve -- the exact numerical solution \re{gap1_solution_exact} of vacuum Eqs.
(\ref{gap2}, \ref{gap1}).
 The dashed curve -- the solution \re{gap1_solution}, obtained by the iterations ($1/N_c$-expansion) with the same accuracy. Data points are from~\cite{Bowman:2005vx}. Notice that the scale of the lattice data is $1.64 \,GeV$, not $\rho^{-1}\approx 0.6\, GeV$. }
 \label{fig:1}
 \end{figure} 
 Fig.\ref{fig:1} represents the $M(m)$-dependence obtained from Eqs.
(\ref{gap2},\ref{gap1}).
 For the sake of comparison we also plotted the  lattice data from~\cite{Bowman:2005vx}.
From the first point of view our results are $\approx 30\%$ higher than the lattice data. However, they are given in different gauges and on different scales. Since $M(p)$ is essentially nonperturbative object, it is not very easy to rescale the data and make comparison. Rough estimates in perturbative QCD show that the discrepancy may be attributed to the scales difference. We may conclude that we have a qualitative correspondence between our model result for $M(m)$-dependence and 
unquenched lattice data ~\cite{Bowman:2005vx}, as it was expected.
 
 \section{Quark condensate}
 \label{SectQuarkCondensate}
  The presence of the quark condensate $\langle\bar qq\rangle$ is one of the most important properties of the QCD vacuum. Its value characterizes the S$\chi$SB. In the chosen framework we can extract it directly from the effective action taking derivative over the current quark mass~\cite{Kim:2005jc}
 \begin{eqnarray}\nonumber
 &&\langle\bar qq\rangle =\frac{1}{2}\frac{\partial \Gamma_{eff}}{\partial m}=
 -\frac{1}{2}\Tr\left(\frac{i}{\hat p +i\mu (p)}-\right.
 \\ 
 &&\left.
 \frac{i}{\hat
 p+im}\right) +\frac{1}{2}\intq V^{\bar qq}(q) \tilde \Pi_i(q),
 \label{cond}
 \end{eqnarray}
 where the first term is a contribution of the tree-level action and the corresponding vertex part $V^{\bar qq}(q)$ is
 \begin{eqnarray}
 \nonumber
 &&V^{\bar qq}(q)=\Tr\left(
  \frac{M F^2(p)}{(\hat p+i\mu(p))^2}\Gamma_i
  \frac{M F^2(p+q)}{\hat p+\hat q+i\mu(p+q)}\Gamma_i\right)
 \end{eqnarray}
 Evaluation of (\ref{cond}) gives 
 \begin{eqnarray}
 -\langle\bar qq\rangle (m) &=&\left(\left( 0.00497 - 0.0343\,m \right) \,N_c+\right.\\
 &&+\left(0.00168 - 0.0494\,m\right. \nonumber\\
 &&\left.\left.- 0.0580\,m\,\ln m\right)\right)[GeV^3]
 +{\cal O}\left(m^2,\frac{1}{N_c^2}\right)\nonumber
 \end{eqnarray}
 \begin{figure}[h]
   \centering
 \includegraphics[scale=0.4]{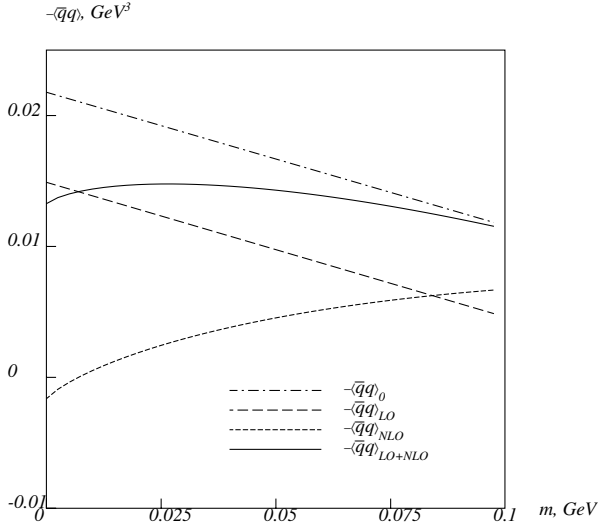}
 \caption{$m$-dependence of the quark condensate $-\langle\bar qq\rangle$.  The long-dashed curve is the LO result $-\ave{\bar qq}_{LO}$, the short-dashed curve is the NLO contribution $-\ave{\bar qq}_{NLO}$, the solid curve is the total contribution $-\ave{\bar qq}_{LO+NLO}$. The dot-dashed line represents the leading-order in $1/N_c$-expansion result, evaluated with the mass $M_0$ from (\ref{gap1_solution}) (see text).}
 \label{fig:2}
 \end{figure}
 The $\langle\bar qq\rangle (m)$-dependence is depicted on the Fig.~\ref{fig:2}. We can see again that due to the chiral logarithm the $m$-dependence is not linear and meson loops change the $m-$dependence of the quark condensate drastically.
  
  Here and below we define as leading order (LO) the result calculated with the $1/N_c$-corrections dropped everywhere except in the mass $M(m)$, which is taken from \re{gap1_solution_exact} without the last ${\cal O}(m/N_c, m/N_c\ln m )$-terms. As next-to-leading order (NLO) we define the contribution of the last two terms in  \re{gap1_solution_exact} plus "direct" contribution of meson loops.
  
  For the sake of comparison, we also plotted the value $\ave{\bar qq}_0$ which one would get using LO formulae with the mass $M_0(m)$ taken from~(\ref{gap1_solution}).
Recall that the value $\ave{\bar qq(m=0)}=(255\,MeV)^3$, as well as $F_\pi(m=0)=88\, MeV$, was used  as the input  in order to fix the parameters $(\rho, R)$ in~(\ref{rhoR}).
In LO we have $\langle\bar qq\rangle (m=0.1)/\langle\bar qq\rangle (m=0)|_{LO}=0.31$, while with NLO corrections we have $\langle\bar qq\rangle (m=0.1)/\langle\bar qq\rangle (m=0)=0.89$, a noticeably different result showing the size of the chiral logarithmic term in the NLO-corrections.

 \section{Quarks in  external axial-vector field and pion properties}
The formalism of the effective action used in one of the previous sections may be successfully applied in the presence of the external axial-vector isovector field $a_\mu=a_{\mu}^{i}\tau_i /2$.
 The general partition function~\re{Z2} is reduced in this case to the form
\begin{eqnarray}
\label{Zma}
Z_N[m,\vec a_\mu]=\int d\lambda \exp(-\Gamma_{eff}[m,\lambda, \vec u, \vec a_\mu])
\end{eqnarray}
The external field $\vec a_\mu$ can generate nonzero vacuum average $\ave{\vec \phi}=\vec u$ and shift the value of the vacuum filed $\ave{\sigma}$ and saddle-point value $\lambda$. 

 In this paper we restrict ourselves to the case of the soft and weak external field $a_\mu(q)$, which can be treated in perturbative fashion, and $q\sim M_\pi \ll \rho^{-1}$. For the purpose of this paper it is sufficient to keep only ${\cal O}(\vec a_\mu^2, \vec a_\mu \partial_\mu\vec u, \partial_\mu\vec u\partial_\mu\vec u)$-terms in (\ref{Zma}).

On general grounds, one can guess that the vacuum expectation value $\vec u\sim \vec a_\mu$, whereas shifts of the vacuum field $\ave{\sigma}$ and saddle-point value $\lambda$ are proportional to the second power of $\vec a_\mu$. Using the saddle-point equation
\begin{eqnarray}
\frac{\partial\Gamma_{eff}[m,\lambda, \vec u, \vec a_\mu]}{\partial\lambda}=0
\end{eqnarray}
and the vacuum equations
\begin{eqnarray}
\frac{\partial\Gamma_{eff}[m,\lambda, \vec u, \vec a_\mu]}{\partial\sigma}=0,\;
\frac{\partial\Gamma_{eff}[m,\lambda, \vec u, \vec a_\mu]}{\partial\vec u}=0,
\label{vacuumSigmau}
\end{eqnarray}

we may easily get that the shifts of $\ave{\sigma}, \lambda$ contribute only to ${\cal O} (a^4)$-terms and thus may be safely omitted in this paper.

In the leading order the effective action simply coincides with the classical $S$ defined in (\ref{S}).
Using vacuum equations~(\ref{vacuumSigmau}), one may show that $\ave{\sigma}^2+\ave{\vec \phi}^2=const$. This inspires us to introduce a unitary matrix $U$ with the properties
\begin{eqnarray}
&&U=u_0+i\vec\tau\vec u,\,\, U^\dagger U=UU^\dagger=1,\\
&&\ave{\sigma}=\sigma u_0,\;\ave{\vec \phi}=\sigma \vec u.
\label{U:definition}
\end{eqnarray}
where $\sigma$ is the value found in Section \ref{SectDynMass}. In this representation the vacuum meson field is represented as $\Phi_{cl}=\sigma\, U$.

In the next to leading order one has to take into account the fluctuations of the field $\Phi\to \sigma U + \Phi'$ and integrate over $\Phi'$.

Meson loop contribution to $\Gamma_{eff}$ has a form
\begin{eqnarray}
&&\Gamma^{mes}_{eff}[m,\lambda, \vec u, \vec a_\mu]=
\frac{1}{2}\Tr\ln \frac{\delta^2  S[m,\lambda,\sigma,\vec u,\vec a_\mu,\Phi']}{\delta\Phi'_i
\delta\Phi'_j}|_{\Phi' =0}\nonumber
\\
&=&
\Gamma^{mes}_{eff}[m,\lambda, \vec u=0, \vec a_\mu=0]+\Delta\Gamma^{mes}_{eff}[m,\lambda, \vec u, \vec a_\mu]
\end{eqnarray}

It is convenient to rewrite the second derivative $\delta^2S/\delta \Phi^{'2}$ as
\begin{eqnarray}
\frac{\delta^2S}{\delta \Phi'_i\delta \Phi'_j}={\cal O}(a^0)+A_{1,ij}+A_{2,ij}+{\cal O}(a^3)
\end{eqnarray}
where $A_{1,ij}$ is of the first order in external field $a$ and induced field $u$, and
$A_{2,ij}$ is of the second order in them.
Then
\begin{eqnarray}
\Delta\Gamma^{mes}=
\frac{1}{2}\Tr\Pi_i\delta_{ij}A_{2,ij} -\frac{1}{4}\Tr[\Pi_i\delta_{ij}A_{1,ij}]^2 
\label{deltagamma}
\end{eqnarray}
Expanding (\ref{deltagamma}) and collecting the terms $a_\mu a_\nu$, $a_\mu \partial_\nu u_i$ and $\partial_\nu u_i\partial_\mu u_j$, after simple but very tedious evaluations it is possible to show that in agreement with chiral symmetry expectations, the structure of the effective action is
\begin{eqnarray}
\label{gamma}
&&\Gamma_{eff}=\alpha_0 (\vec a_\mu+\partial_\mu \vec u)^2+m\,\alpha_1 \partial_\mu \vec u (\vec a_\mu+ \partial_\mu \vec u)+\\
&&+m\, \alpha_2 \vec u^2=\frac{1}{2}\left[F_{aa}^2 \vec a_\mu^2+F_{uu}^2  (\partial_\mu \vec u)^2+2 F_{au}^2 \vec a_\mu\partial_\mu \vec u +\nonumber\right.\\
&&\left.+ F_{uu}^2  M_\pi^2 \vec u^2\right]+{\cal O}(a^3,u^3, m^2),\nonumber
\end{eqnarray}
where the constants $F_{ij}$ differ only beyond chiral limit:
\begin{eqnarray}
F_{aa}^2 -F_{uu}^2=2\left(F_{au}^2 -F_{uu}^2\right)=-\alpha_1\,m.
\label{Fpi:Differences}
\end{eqnarray}
From (\ref{gamma},\ref{Fpi:Differences}) one can get that the two-point axial-isovector currents correlator has a form:
\begin{eqnarray}
\label{aa:structure}
&&\int d^4 x e^{-iq\cdot x}\langle j^{A,i}_\mu(x)j^{A,j}_\nu(0)\rangle=\\
&&=\delta_{ij}F_\pi^2\left(\delta_{\mu\nu}-\frac{q_\mu q_\nu}{q^2+M_\pi^2}\right)+{\cal O} (q^2)\nonumber
\end{eqnarray}
We can see that  $M_\pi$ has a meaning of pion mass and $F_\pi$ -- pion decay constant.
Numerically it is much easier to calculate $F^2_\pi$ as a constant in front of $\delta_{\mu\nu}$-term in (\ref{aa:structure}), taking $\vec u(x)=0$, $a_\mu(x)=const$ (this corresponds to $a_\mu(q\approx 0)$). Also, it is possible to show that the result of such evaluation is independent of the path choice in the transporter $L$ (see Section \ref{SectPathIndep} for more details). In a similar way, we can put $a=0, u=u(q)$ and evaluate in NLO the quantities $F^2_{uu}$ and pion mass $M_\pi$.
 Both quantities $F_\pi$ and $M_\pi$ naturally have the chiral log terms due to the pion loops contributions. The coefficients in these chiral log terms are controlled by the low-energy theorems~\cite{LangackerPagels73} and are reproduced analytically in Section~\ref{SectChiralLogs}.

 We can see that the specific structure of the  $\Gamma_{eff}$ \re{gamma} provides a check of the numerical calculations. Moreover, the chiral log theorems provide another check of the numerical calculations.

\subsection{Pion decay constant $F_\pi$ from $a^2$-term}
\label{SectAAcorrelator}
\begin{figure}[h]
  \centering
\includegraphics[scale=0.2]{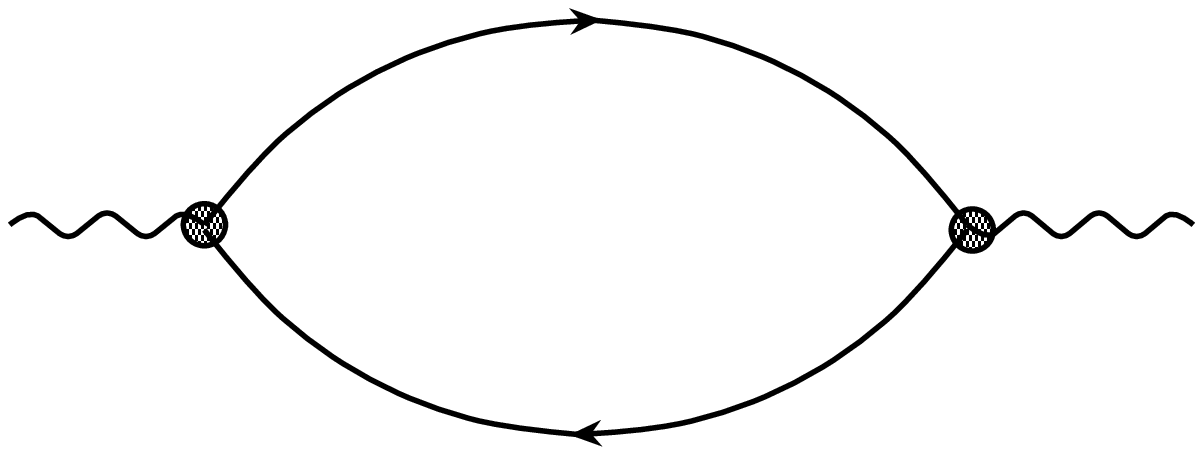}
\includegraphics[scale=0.4]{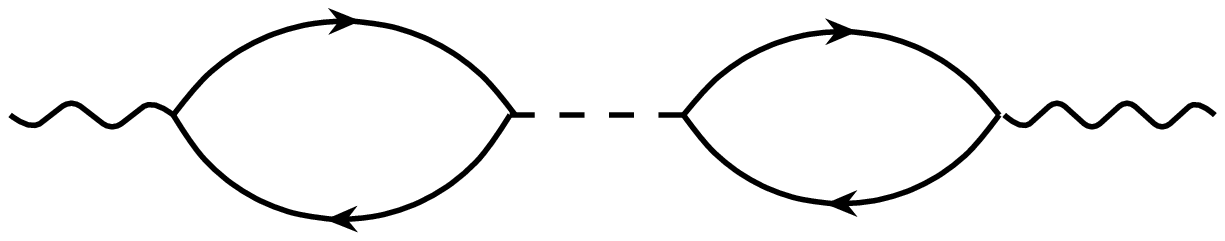}
\caption{The basic diagrams which contribute to the $a^2$ term in $\Gamma_{eff}$. The wavy line corresponds to the external field $a_\mu(x)$, the dashed line corresponds to the intermediate meson, the bulbs correspond to all the possible (local and nonlocal) couplings of the field $a$ to the constituent quarks.}\label{fig:aa}
\end{figure}

\begin{figure}[h]
  \centering
\includegraphics[scale=0.2]{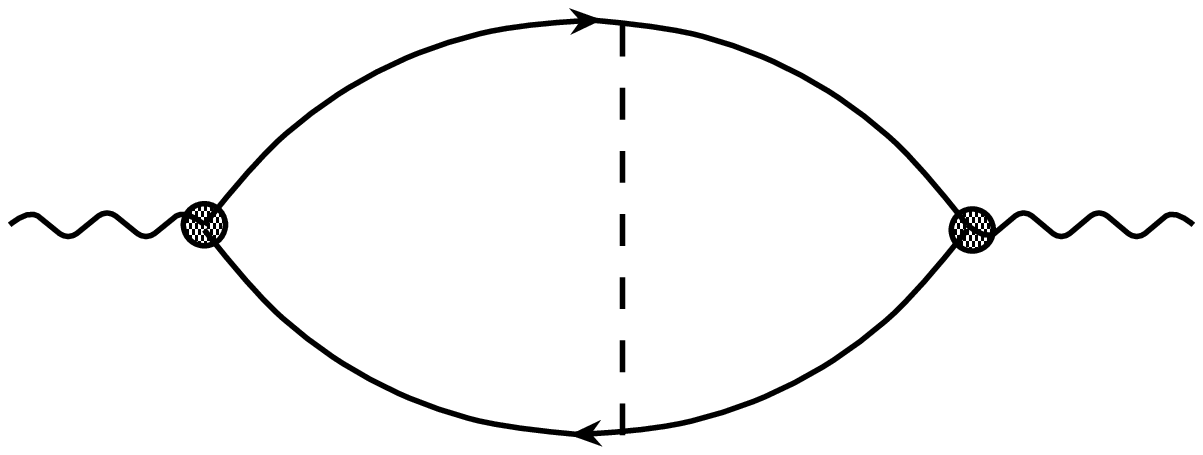}
\includegraphics[scale=0.2]{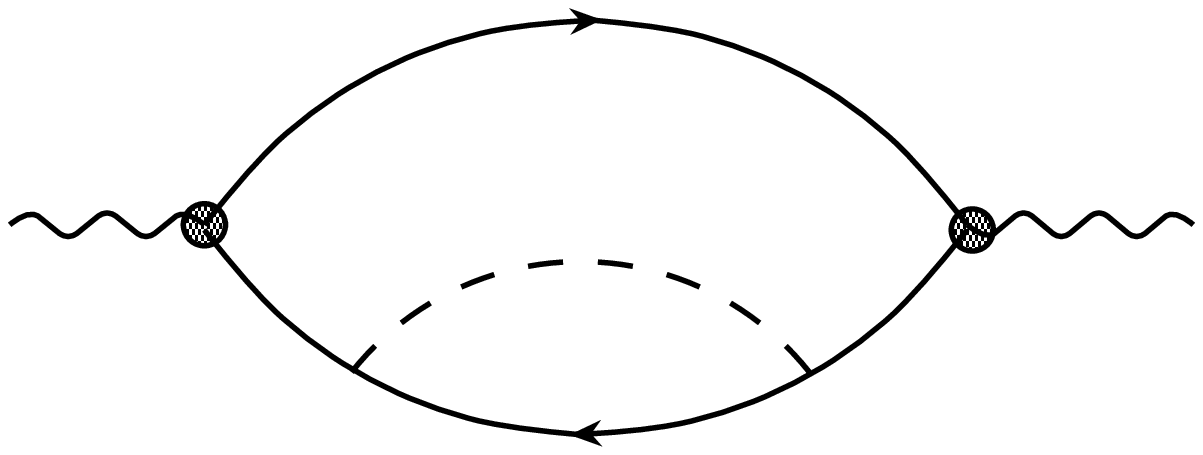}
\includegraphics[scale=0.2]{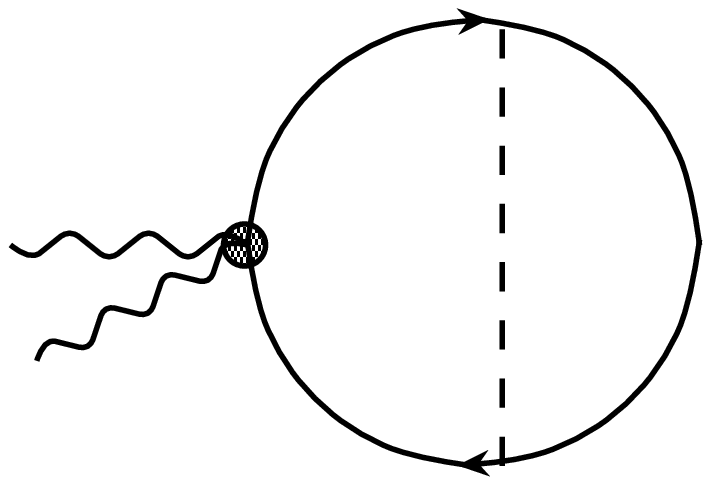}
\includegraphics[scale=0.2]{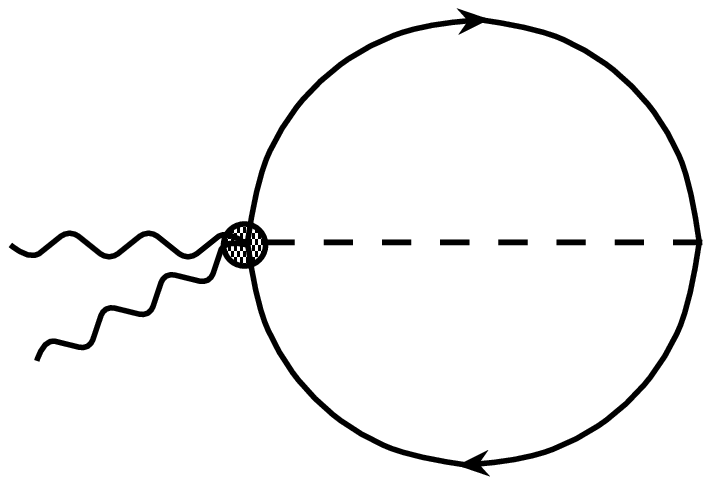}
\includegraphics[scale=0.2]{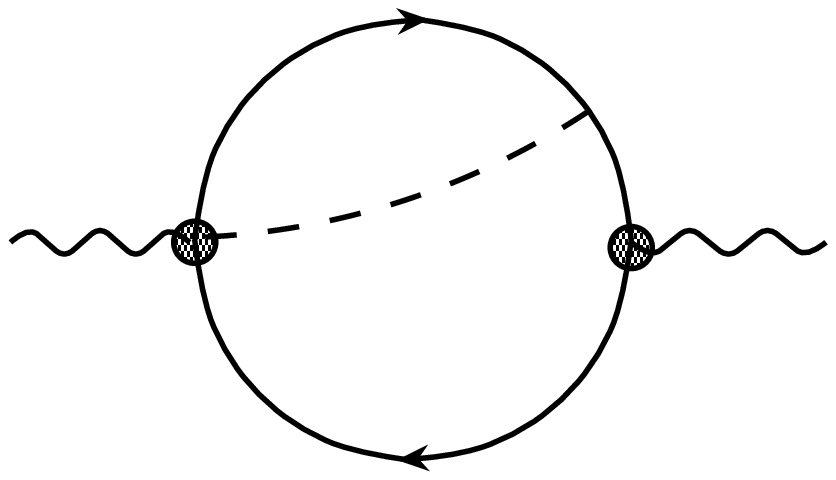}
\includegraphics[scale=0.3]{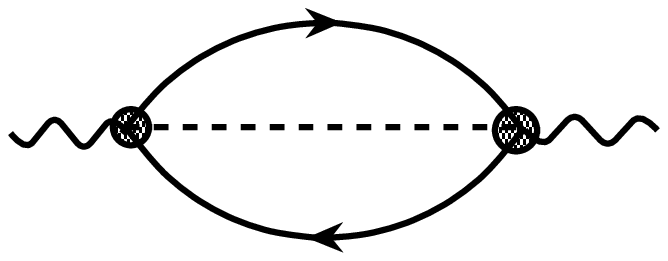}\\
\includegraphics[scale=0.2]{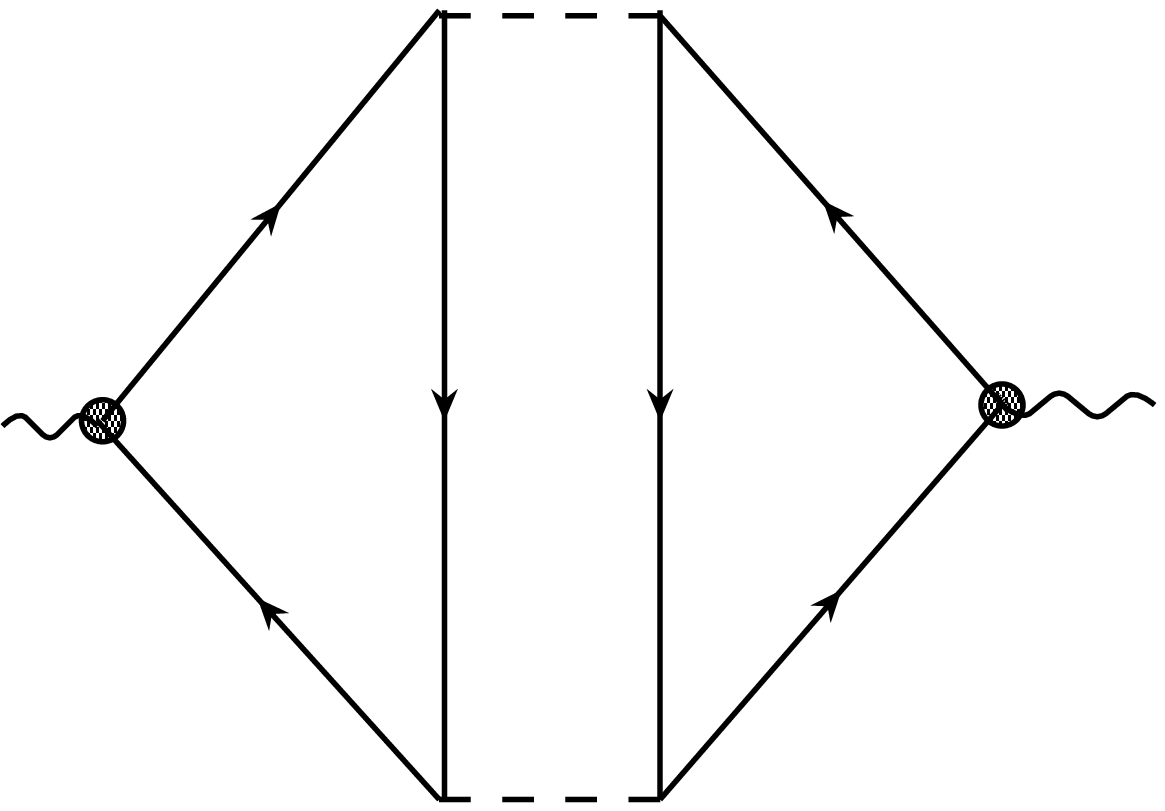}
\includegraphics[scale=0.2]{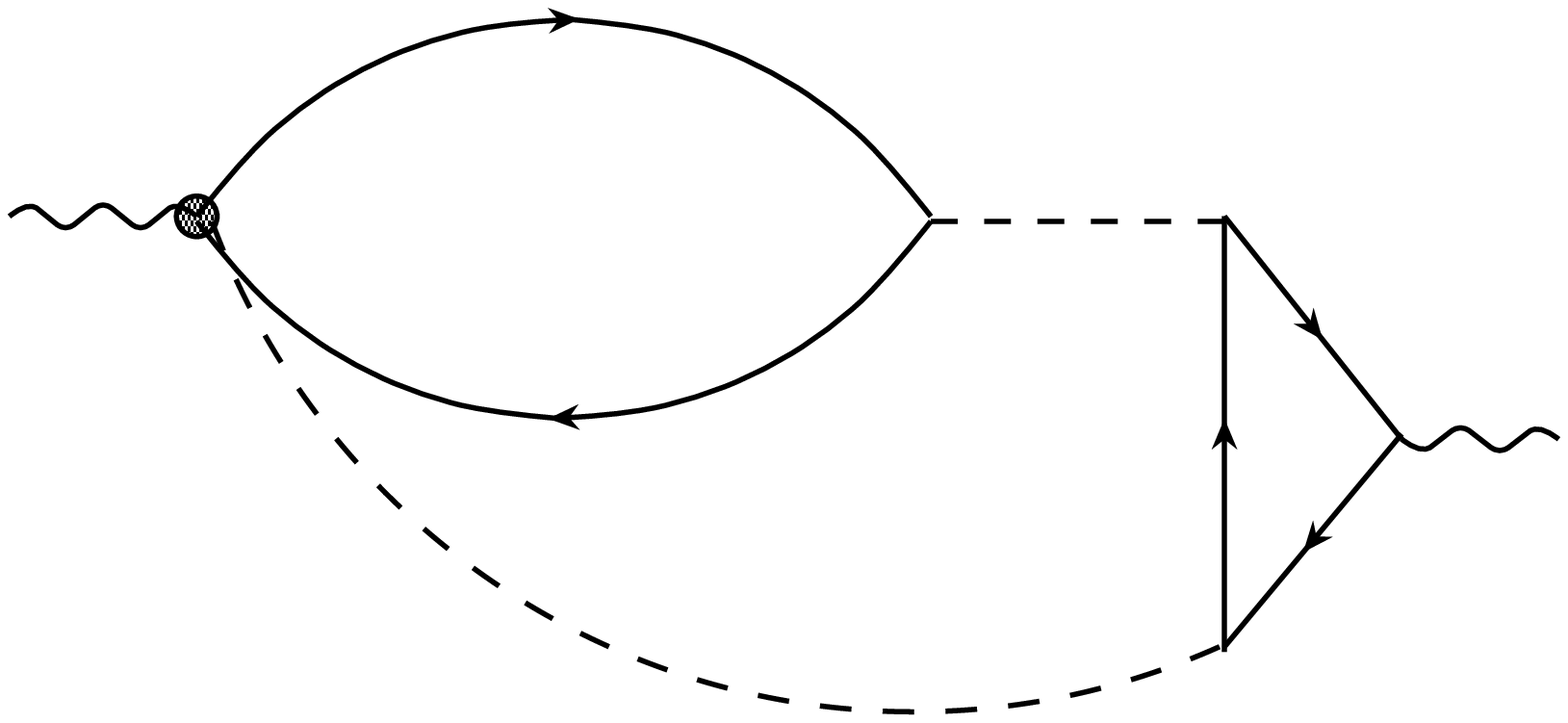}
\caption{The meson loop corrections to the $a^2$ term. The notations are the same as in Fig.~\ref{fig:aa}. }\label{fig:aa-NLO}
\end{figure}

The basic diagrams which contribute to this quantity in the leading order and in the next-to-leading order are shown schematically in the Fig.~\ref{fig:aa} and Fig.~\ref{fig:aa-NLO} respectively. Notice that in integration over $\vec \phi$ the saddle-point is shifted to $\langle\vec \phi\rangle\sim \vec a_\mu$, where the ''proportionality'' sign implies some nonlocal linear operator. All the vertices on these plots should be understood as a sum of the local and nonlocal parts, as it is explained in the Appendix~\ref{SectVertices} and in the Fig.~\ref{fig:Vertices}. 
\par
\begin{figure}[h]
\centering
\includegraphics[scale=0.4]{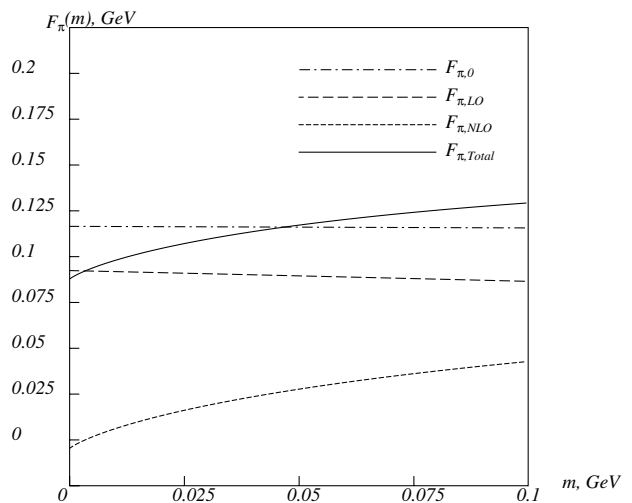}
\caption{$m$-dependence of the pion decay constant $F_\pi$.  The long-dashed curve is the LO contribution, the short-dashed curve is the NLO contribution, the solid curve is the total LO+NLO contribution.  The dot-dashed line represents the leading-order in $1/N_c$-expansion result, evaluated with the mass $M_0$ from (\ref{gap1_solution}) (see text). \label{fig:Fpi(m)}}
\end{figure}

 Finally we have 
\begin{eqnarray}
\label{Res:Fpi}
&&F_\pi^2=N_c\left(\left(2.85 -\frac{0.869}{N_c}\right)-\left(3.51+\frac{0.815}{N_c}\right)m-\nonumber\right.\\
&&\left.-\frac{44.25}{N_c}\,m\,\ln m +{\cal O} (m^2)\right)\cdot 10^{-3}\; [GeV^2]=\\
&&\left(7.67 - 11.35\,m - 44.25\,m\,\ln m\right)\cdot 10^{-3}\; [GeV^2]\nonumber
\end{eqnarray}
$m$ is given in $GeV$ and the constant in front of ${\cal O} (m^0)$ is given in $GeV^2$, the constant in front of ${\cal O} (m)$ is given in $GeV$. The $F_\pi(m)$-dependence is shown in the Fig.\ref{fig:Fpi(m)}.
For the sake of comparison, we also plotted the value $F_{\pi,0}$ which one would get using LO formulae with the mass $M_0(m)$ taken from~(\ref{gap1_solution}).
Recall that the value $F_\pi(m=0)=88\, MeV$, as well as $\ave{\bar qq(m=0)}=(255\,MeV)^3$, was used  as the input  in order to fix the parameters $(\rho, R)$ in~(\ref{rhoR}). The comparison between the solid curve and the long-dashed one shows that the effect of the NLO-corrections grows with $m$ and is about $40\%$ at $m=0.1\,GeV$.

\subsection{$M_\pi$ from pion propagator}
Effective action $\Gamma_{eff}$, Eq.~\re{gamma} at $a=0$ has a meaning of inverse  $\pi$-meson propagator at small external momentum $q\sim M_\pi$ with account of meson loops. For our purpose it is sufficient to have only the ${\cal O}(q^0)$ and ${\cal O}(q^2)$ terms.

The LO propagators of the mesons were defined in (\ref{Meson_Propagators}).
The inverse pion propagator with account of meson loops  is given by
\begin{widetext}
\begin{eqnarray}
&&\Pi^{(ab)-1}_\phi(q)=
\label{propagator_total}\left[4\delta^{ab}+\frac{1}{\sigma^2} Tr_p\left(Q(p)i\gf \tau^aQ(p+q)i\gf \tau^b \right)\right]+\nonumber\\
&&+\frac{1}{\sigma^4}\intk \Pi_{ij}(k)V^{\phi}_{ij}(k,q)-\frac{4}{\sigma^6}\intk \Pi_{i}(k)\Pi_{j}(k+q)V^\phi_i(k,q)V^\phi_j(k,q)
\end{eqnarray}
where the function $Q(p)$ was defined in (\ref{Q(p)}),
and the vertices are
\begin{eqnarray}
&&V^{\phi}_{ij}(k,q)=2 \Tr\left(Q(p)i\gf \tau^aQ(p+q)\Gamma_iQ(p+q+k)\Gamma_jQ(p+q)i\gf\tau^b\right)+ \\\nonumber
&&\Tr\left(Q(p)i\gf \tau^aQ(p+q)\Gamma_iQ(p+q+k)i\gf\tau^bQ(p+k)\Gamma_j\right)\\
&&V^\phi_i(k,q)=\Tr\left(Q(p)i\gf\tau^aQ(p+q)\Gamma_iQ(p+q+k)\Gamma_j\right)
\end{eqnarray}
\begin{figure}[h]
\includegraphics[scale=0.4]{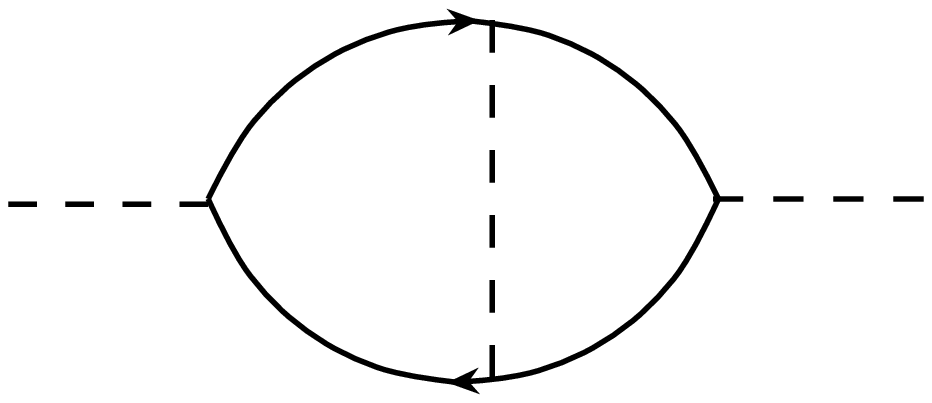}
\includegraphics[scale=0.4]{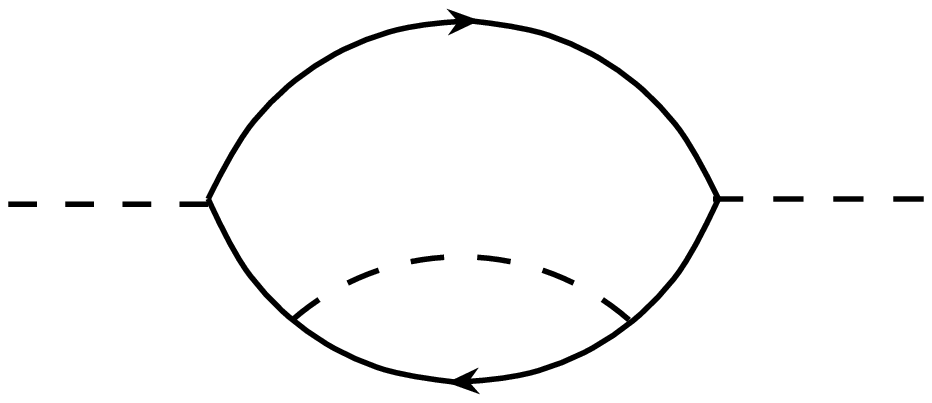}
\includegraphics[scale=0.4]{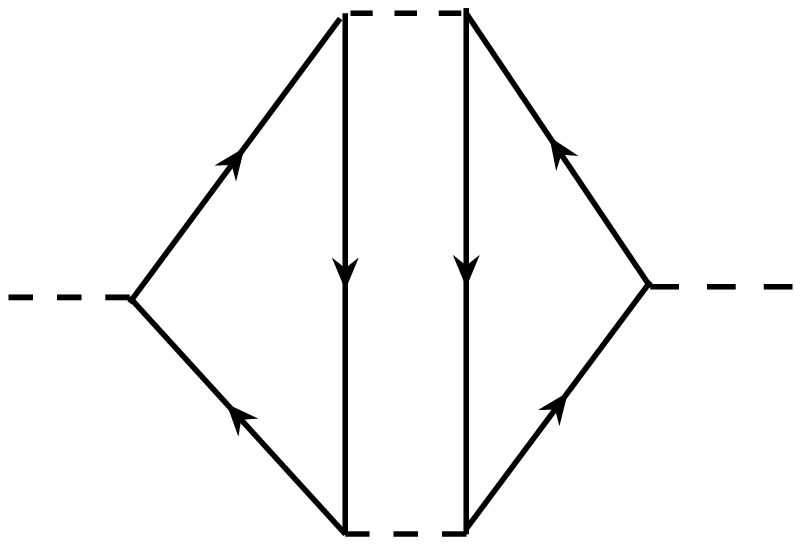}
\caption{\label{Fig2prop}Diagrams corresponding to the last two meson loop terms in (\ref{propagator_total})}
\end{figure}
\end{widetext}
\par
The value of the pion mass $M_\pi$ is defined as a pole position in the propagator (\ref{propagator_total}).
Now we would like to demonstrate that $M_\pi^2\sim m$, i.e. that the loop corrections do not violate the Goldstone theorem, i.e. $\Pi^{-1}_\phi(0)~\sim m$. 
\par
In the leading order over $N_c$ the only contribution to $\Pi^{-1}_\phi(q)$ comes from the first term in (\ref{propagator_total}), and using equation (\ref{gap2leading}) may be reduced to
\begin{eqnarray}
&&\tilde\Pi_{\phi}^{-1}(0)=m\frac{8N_c}{\sigma^2}\intp \frac{M F^2(p)}{p^2+\mu^2(p)} 
\label{bcprop}\end{eqnarray} 
In the next-to-leading order contribution of the first term in (\ref{propagator_total}) is
\begin{widetext}
\begin{eqnarray}
&&\frac{1}{\sigma^4}\intq \tilde\Pi_i(q)V_3^i(q)+
m\frac{8N_c}{\sigma^2}\frac{M_1}{M_0}M \frac{\partial}{\partial M} \left(\intp \frac{M F^2(p)}{p^2+\mu^2(p)}\right)\\ \nonumber 
&&
=\frac{1}{\sigma^4}\intq \tilde\Pi_i(q)V_3^i(q)+
m\frac{8N_c}{\sigma^2}\frac{M_1}{M_0}\left(\intp \frac{M F^2(p)(p^2-M^2 f^4(p)+m^2)}{(p^2+\mu^2(p))^2}\right)
\end{eqnarray}
and explicit expression for $V_3^i(q)$ defined in (\ref{V3def}) is
\begin{eqnarray}
&&V_3^i(q)=8 N_c c_i\intp\left[\frac{ M^3F^4(p)F^2(p+q)\left( \pm 2\, \left( p^2 + p\cdot k\right) \mu(p) -
p^2\mu(p + q) + \mu^2(p)\mu(p + q) \right)}{\left( p^2 + \mu^2(p) \right)^2\left(\left( p + q \right)^2 + \mu(p + q)^2 \right)} \right]\nonumber
\end{eqnarray}
\end{widetext}
where here and below
\begin{eqnarray} 
c_i=\left\{\ba{1, i=\sigma, \eta}\\ {-3, i=\vec \sigma, \vec\phi}\ea\right.
\end{eqnarray}
and the sign in the numerator of the first term is $+$ for $\eta, \phi$ and $-$ for
$\sigma, \vec \sigma$
\par The vertex part $V^\phi_i(k,q=0)$ in the second term of (\ref{propagator_total}) may be rewritten explicitly as
\begin{widetext}
\begin{eqnarray}\label{pcontr2}
&&V_i^\phi(k)=\\
&&\nonumber-8 N_c \intp\left[2 c_i\frac{M^4F^6(p)F^2(p+k)(\mu(p)\mu(p+k)\pm (p^2+p\cdot k) )}{(p^2+\mu^2(p))^2((p+k)^2+\mu^2(p+k))}
+\frac{M^2F^4(p)}{p^2+\mu^2(p)}\frac{M^2F^4(p+k)}{(p+k)^2+\mu^2(p+k)}\right]
\end{eqnarray}
\end{widetext}
\par The third term in (\ref{propagator_total}) may be reduced to the form
\begin{eqnarray}
-\frac{4}{\sigma^6}\intk (\tilde\Pi_\sigma(k)\tilde\Pi_{\vec\phi}(k)+\tilde\Pi_{\eta}(k)\tilde\Pi_{\vec\sigma}(k))\bar V^2(k)
\label{pcontr3}\end{eqnarray}
where $\bar V(k)$ equals
\begin{eqnarray}
\hspace{-3mm}\bar V(k)=8N_c \intp\frac{M^2F^4(p)}{p^2+\mu^2(p)}\frac{M\, F^2(p+k)\mu(p+k)}{(p+k)^2+\mu^2(p+k)} \label{VertexV}
\end{eqnarray}
To eliminate the terms with double propagators in the third term, we follow the strategy used in \cite{Dmitrasinovic:1995cb} for pure NJL. 
Using explicit expressions for propagators, we can notice that
\begin{eqnarray}
&&\tilde\Pi^{-1}_{\sigma}(q)-\tilde\Pi^{-1}_{\vec\phi}(q)=\tilde\Pi^{-1}_{\eta}(q)-\tilde\Pi^{-1}_{\vec\sigma}(q)=\\ \nonumber
&&\hspace{-3mm}\frac{16 N_c}{\sigma^2}\intp\frac{MF^2(p)\mu(p)}{p^2+\mu^2(p)}\frac{MF^2(p+q)\mu(p+q)}{(p+q)^2+\mu^2(p+q)} =\frac{2 V(q)}{\sigma^2}
\label{Propagators:relation}
\end{eqnarray}
where
\begin{eqnarray}
\hspace{-3mm}V(q)\equiv 8 N_c\intp\frac{MF^2(p)\mu(p)}{p^2+\mu^2(p)}\frac{MF^2(p+q)\mu(p+q)}{(p+q)^2+\mu^2(p+q)}
\end{eqnarray}
Obviously
\begin{eqnarray}
&&\bar V(q)=V(q)+m\, \delta V(q)\\
&&\delta V(q)\equiv
-8 N_c\intp\frac{MF^2(p)}{p^2+\mu^2(p)}\frac{MF^2(p+q)\mu(p+q)}{(p+q)^2+\mu^2(p+q)}\nonumber
\end{eqnarray}
Hence
\begin{eqnarray}
\tilde\Pi_\sigma(q)\tilde\Pi_{\vec\phi}(q)=-\sigma^2\frac{\tilde\Pi_{\sigma}(q)-\tilde\Pi_{\vec\phi}(q)}{2 V(q)} \\
\tilde\Pi_\eta(q)\tilde\Pi_{\vec\sigma}(q)=-\sigma^2\frac{\tilde\Pi_{\eta}(q)-\tilde\Pi_{\vec\sigma}(q)}{2 V(q)}
\end{eqnarray}
and (\ref{pcontr3}) turns into
\begin{eqnarray}
&&\frac{2}{\sigma^4}\intk (\tilde\Pi_\sigma(k)+\tilde\Pi_{\eta}(k)-\tilde\Pi_{\vec\sigma}(k)-\tilde\Pi_{\vec\phi}(k))V(k)\times\nonumber\\ 
&&\left(1+2\, m\, \frac{\delta V(k)}{V(k)}+m^2\left(\frac{\delta V(k)}{V(k)}\right)^2\right)
\label{pcontrtot2}
\end{eqnarray}
We can notice that coefficients (signs) in front of each of the terms in (\ref{pcontrtot2}) may be rewritten as
$\frac{c_i+1}{2}$. Finally after several simplifications we arrive to
\begin{widetext}
\begin{eqnarray}\label{FinalPi}
&&m \frac{8N_c}{\sigma^4} \intk
\tilde\Pi_i(k)\intp\left[2 c_i\frac{(\mu(p)\mu(p+k)\pm \left( p^2 + p\cdot k\right))M^3F^4(p)F^2(p+k) }{\left( p^2 + \mu^2(p) \right)^2\left(\left( p + k \right)^2 + \mu(p + k)^2 \right)}
+\right. \label{pcontrtotfinal2}\\   \nonumber
&&\left. 
-\frac{M\, F^2(p)}{p^2+\mu^2(p)}\frac{M\, F^2(p+k)}{(p+k)^2+\mu^2(p+k)}\left(m+c_i\mu(p)\right)\right] +m^2\frac{(c_i+1)}{\sigma^4}\intk \tilde\Pi_i(k)\frac{\delta V^2(k)}{V(k)} \nonumber\\
&&+
m\frac{8N_c}{\sigma^2}\frac{M_1}{M_0}\left(\intp \frac{M F^2(p)(p^2-M^2 f^4(p)+m^2)}{(p^2+\mu^2(p))^2}\right)
\nonumber
\end{eqnarray}
\end{widetext}
We can see that the pion propagator $\Pi_\phi^{-1}(0)\sim m$, i.e. the loop corrections do not violate the Goldstone theorem. Notice that the function $V(k)$ is positively defined, so no unexpected poles should appear due to the term $\propto\intk \Pi_i(k)\frac{\delta V^2(k)}{V(k)}$. 
\par
\begin{figure}[h]
\includegraphics[scale=0.4]{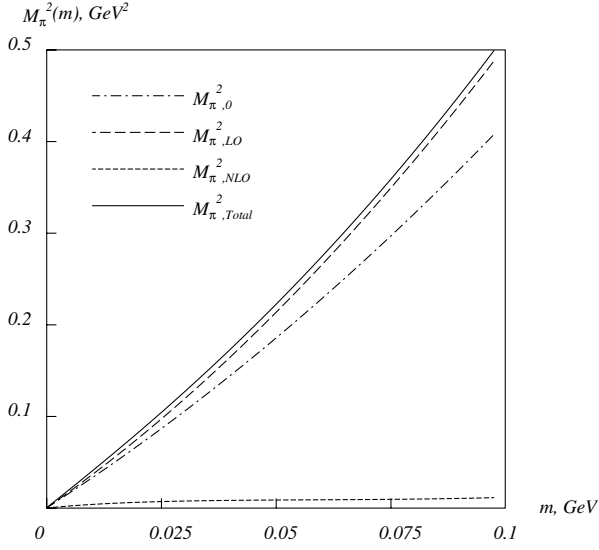}
\caption{$m$-dependence of the pion mass  $M_\pi$.  The long-dashed curve is the LO contribution, the short-dashed curve is the NLO contribution, the solid curve is the total LO+NLO contribution. The dot-dashed line represents the leading-order in $1/N_c$-expansion result, evaluated with the mass $M_0$ from (\ref{gap1_solution}) (see text).\label{fig:4}}
\end{figure}

 Evaluating $F^2_{uu}$-term in (\ref{gamma}) with account of meson loops and taking the ratio of~(\ref{FinalPi}) to $F^2_{uu}$, we obtain the pion mass
\begin{eqnarray}
\label{Res:Mpi}
&&M_\pi^2=m\left(\left(3.49+\frac{1.63}{N_c}\right)+\right.\\\nonumber
&&\left.m\left(15.5+\frac{18.25}{N_c}+\frac{13.5577}{N_c} \ln m \right)+{\cal O}(m^2)\right)=\\
&&m\,(4.04\,+21.587\,m+4.52\,m\ln m +{\cal O}(m^2))[GeV^2]\nonumber
\end{eqnarray}
in the next-to leading order. The $M_\pi(m)$-dependence of the pion mass is shown on the Fig.\ref{fig:4}.
For the sake of comparison, we also plotted the value $M_{\pi,0}$ which one would get using LO formulae with the mass $M_0(m)$ taken from~(\ref{gap1_solution}). Altogether, for this observables the NLO-corrections turn out to be small.
 
\section{Finite width corrections}
\label{SectFWC}
In the previous sections it was assumed that the instanton size distribution has a zero width, i.e. 
\begin{eqnarray}
d(\rho)=\delta(\rho-\bar\rho)
\end{eqnarray}
and all instantons have the same size $\bar\rho$. As it was shown in the early works of Diakonov, Petrov\cite{Diakonov:1983hh,Diakonov:1985eg}, this approximation is justified by the small parameter $1/N_c$, i.e.
\begin{eqnarray}
&&\frac{\ave{\rho^2}-\ave{\rho}^2}{\ave{\rho}^2}\sim {\cal O}\left(\frac{1}{N_c}\right).
\label{SizeDefinition}
\end{eqnarray}
For numerical evaluations we take the value
\begin{eqnarray}
\delta\rho^2=\ave{\rho^2}-\ave{\rho}^2\approx\frac{0.5599\, GeV^{-2}}{N_c}
\label{SizeNumber}
\end{eqnarray}
which follows from the two-loop size distribution of \cite{Diakonov:1983hh}
 Since we are interested in all $1/N_c$ corrections, we must take the finite width into account. To do this, we must return to the formula (\ref{tildeV}). Additional integration over $\rho$ doesn't change the exponentiation procedure, and we get the standard $2N_f$-interaction term in the effective action $S$. However, for bosonization we should slightly modify the standard procedure.  Here we consider only the case $N_f=2$ we are mainly interested in. For this case
\begin{eqnarray}
&&e^{-\intzr J^2(z,\rho)}=\\
&&\int D\Phi(z,\rho)\exp\left(-\frac{1}{4}\intzr \Phi^2(z,\rho)+\right.\nonumber\\
&&\left.\intzr \Phi(z,\rho)J(z,\rho)\right)\nonumber
\end{eqnarray}
where $\Phi(z,\rho)$ are pure mathematical objects (matrices). Thus the effective action turns into
\begin{eqnarray}
&&S=\frac{N}{V}\ln \lambda+2\intzr \Phi^2(z,\rho)-\label{EffActSize}\\
&&Tr\ln\left(\hat p+i m+i c\, \lambda^{0.5}\intzr \hat K(z,\rho)\Phi(z,\rho)\right)\nonumber
\end{eqnarray}
where $c$ is some inessential constant and $\hat K_{x,y}(z,\rho)\sim \bar \phi(x-z,\rho)\phi(z-y,\rho)$.
The LO vacuum equations which follow from (\ref{EffActSize}) are
\begin{eqnarray}
&&\frac{N}{V}=\half  Tr\left(\frac{i c\, \lambda^{0.5}\intzr \Phi(\rho)\hat K(z,\rho)}{\hat p+i m +i c\, \lambda^{0.5}\intzr \hat K(z,\rho)\Phi(\rho)}\right)\label{Gap:FWC}\\
&&4\Phi_0(\rho)=Tr\left(\frac{i c\, \lambda^{0.5} \hat K(z,\rho)}{\hat p+i m +i c\, \lambda^{0.5}\intzr \hat K(z,\rho)\Phi_0(\rho)}\right)\nonumber
\end{eqnarray}
where $\Phi_0(\rho)$ has the quantum numbers of $\sigma$. The unknown variables in these equations are parameter $\lambda$ and function $\Phi(\rho)$. In general case these equations are very complicated and may be solved only numerically. However, for the case of small width we can make expansion over $\delta\rho^2=\ave{\rho^2}-\ave{\rho}^2$. In particular, we expand the functions $\Phi(\rho)$ and $\hat K(\rho)$ over $(\rho-\bar\rho)$, get a system of equations for the Taylor coefficients $\Phi_i$ and $\lambda$. Solving these equations, we can restore the function $\Phi(\rho)$ (at least in the small vicinity of $\bar\rho$), and after that evaluate the finite width contributions to all the quantities. Details of these evaluations are given in Appendix~\ref{SectFWCDetails}. The final results for the corrections are
\begin{eqnarray}
&&\frac{\delta F_\pi^2}{\delta\rho^2}=\frac{N_c}{3}\times\\
&&\nonumber\left(0.0024 - 0.020 m + 0.019 m^2\right)[GeV^4]\\
&&\frac{\delta \ave{\bar qq}}{\delta\rho^2}=\frac{N_c}{3}\times\\
&&\nonumber\left(-0.0024 + 0.058 m - 0.33 m^2\right)[GeV^5]
\end{eqnarray}
Substituting the numerical value (\ref{SizeNumber}), we get
\begin{eqnarray}
&&\delta F_\pi^2=\\
&&\nonumber\left(0.00045 - 0.0037 m + 0.0036 m^2\right)[GeV^2]\\
&&\delta \ave{\bar qq}=\\
&&\nonumber\left(-0.00045 + 0.011 m - 0.062 m^2\right)[GeV^3]
\end{eqnarray}
Here $m$ is given in $GeV.$
Thus we can see that these corrections are relatively small, $\approx 5\%$ for $F_\pi^2$ and $\approx 2.6\%$ for $\ave{\bar qq}$.
\par
\begin{figure}[h]
\includegraphics[scale=0.4]{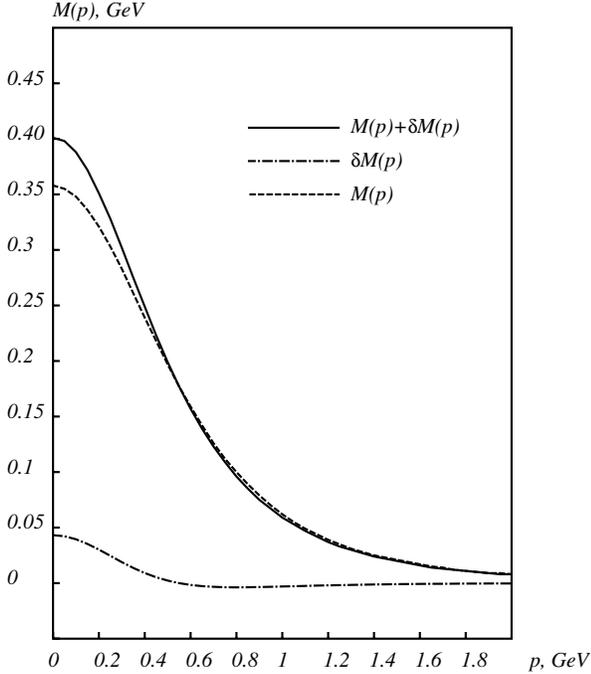}
\caption{Change of the $p$-dependence of the constituent quark mass  $M(p)$ due to the finite width corrections.  The dashed curve is the contribution of the leading order result, the dot-dashed curve is the contribution of the finite width correction, the solid curve is a total result}\label{fig:MFWC(p)}
\end{figure}

 If we define the mass $M(p)$ of the constituent quark as 
\begin{eqnarray}
&&M(p)=c\, \lambda^{0.5}\intzr \hat K(z,\rho)\Phi_0(\rho)\label{MassDefinition:FWC},\\
&&M(p)=M F^2(p)+\delta \rho^2 \delta M_2(p)+{\cal O}\left(\delta\rho^4\right)
\end{eqnarray}
we can get the $p$-dependence shown in the Fig.~\ref{fig:MFWC(p)}. We can see that for $p=0$ the increase of the dynamical quark mass is $\delta M/M=10\%$.
\section{Tensor terms contributions}
\label{SectTensorTerm}
In this section we study contribution of the ($1/N_c$-suppressed) tensor terms to the axial currents correlator $\ave{a_\mu a_\nu}$ introduced in~(\ref{Z:withTensorTerms}). 
 Notice that due to the identity (\ref{gf}) the fields $J^{1,5}_{\mu\nu}$ depend on each other. Under Lorentz transformations $J_{\mu,\nu}$ transforms like $(1,0)+(0,1)$ whereas chiral components $J^{\pm}_{\mu,\nu}$ transform like $(1,0)$ and $(0,1)$.\\
 In case $N_f=2$ evaluation of the determinant becomes especially simple:
\begin{eqnarray}
&&\det J^\pm=\frac{1}{4}\sum_i (J^{(i)\pm})^2=\frac{1}{16}\left(\sum_i (J_i)^2+(J^5_i)^2\right),\\
&& \det J_{\mu\nu}^\pm=\frac{1}{8}\left(\sum_i(J^{(i)}_{\mu\nu})^2\pm\frac{i}{2}\epsilon_{\alpha\beta\mu\nu} J^{(i)}_{\mu\nu}J^{(i)}_{\alpha\beta}\right)\\
&&J^{(i)}=\psi^{\dagger '}\Gamma_i\psi',\;\;J^{5,(i)}=\psi^{\dagger '}\Gamma_i\gf\psi',\\
&&J^{(i)}_{\mu\nu}=\psi^{\dagger '} \sigma_{\mu\nu}\Gamma_i\psi'\\
&&\Gamma_i=\{1,i\vec\tau\}
\end{eqnarray}
The term $\pm\frac{i}{2}\epsilon_{\alpha\beta\mu\nu} J^{(i)}_{\mu\nu}J^{(i)}_{\alpha\beta}$ effectively averages to zero if the vacuum is symmetric, $N_+=N_-$.
For bosonization we use standard formula
\[e^{\alpha^2 x^2/4}\sim\int\, dy e^{-y^2-\alpha x y}\]
Notice however one subtle issue which exists for the tensor fields. The field $J_{\mu\nu}$ is antisymmetric, $J_{\nu\mu}=-J_{\mu\nu}$, thus not all the components are independent and this fact must be taken into account in bosonization to avoid double counting :
\begin{eqnarray}
&&J_{\mu\nu}^2\equiv \sum_{\mu,\nu}J_{\mu\nu}^2=2 \sum_{\mu<\nu}J_{\mu\nu}^2\\
&&\exp\left(\alpha^2 J_{\mu\nu}^2/8\right)=\\
&&\nonumber\int D\Phi_{\mu\nu}\exp\left(-\sum_{\mu<\nu}\Phi_{\mu\nu}^2-\alpha\sum_{\mu<\nu}J_{\mu\nu}\Phi_{\mu\nu}\right)=\label{BosonisationTensor}\\
&&=\int D\Phi_{\mu\nu} \exp\left(-\half\Phi_{\mu\nu}^2-\half\alpha J_{\mu\nu}\Phi_{\mu\nu}\right)\nonumber
\end{eqnarray}
The above-mentioned antisymmetry must be also taken into account in all further differentiations over $\Phi_{\mu\nu}$.\\
From (\ref{BosonisationTensor}) we  get for the partition function
\begin{eqnarray}
&&Z_N=\int d\lambda_+d\lambda_-D\bar\psi D\psi D \Phi^\pm D \Phi_{\mu\nu}^\pm e^{-S}\\
&&S=-N_\pm \ln\lambda_\pm+2\left(\Phi_i^2+\half\Phi_{i,\mu\nu}^2\right)+\\
&&\nonumber\psi^\dagger \left[i\hat \partial+\hat v +im + \right.\\ \nonumber
&&\left. i \lambda^{0.5}\bar L F(p)\left(\alpha \Phi_i\Gamma_i+\half\beta \Phi_{i,\mu\nu}\sigma_{\mu\nu}\Gamma_i\right) F(p) L^{-1}\right]\psi
\end{eqnarray}
We can integrate out fermions and get
\begin{eqnarray}
&&Z_N[v,T]=\int d\lambda d\lambda D \Phi D \Phi_{\mu\nu} e^{-S}\\
&&S=-N_\pm \ln\lambda_\pm+2\left(\Phi_i^2+\half\Phi_{i,\mu\nu}^2\right)-\label{EffAction}\\
&&\nonumber Tr\log \left[\hat p+\hat v+\frac{i}{2}T_{\mu\nu}\sigma_{\mu\nu}+im + \right. \\
&&\left. i \lambda^{0.5}\bar L F(p)\left(\alpha \Phi_i\Gamma_i+\half\beta \Phi_{i,\mu\nu}\sigma_{\mu\nu}\Gamma_i\right) F(p) L^{-1}\right]\nonumber
\end{eqnarray}
where we have introduced external \textit{tensor} source $T_{\mu\nu}$. 

\subsection{Contribution to the axial currents correlator}
\begin{figure}
\includegraphics[scale=0.5]{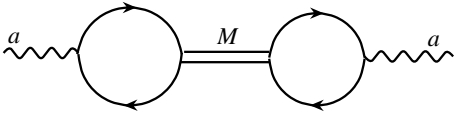}
\caption{\label{Fig:aTa} Contribution to the axial currents correlator. The intermediate state is the tensor meson $\Phi_{\mu\nu}$, the diagram is $1/N_c$-correction.}
\end{figure}

In this section we consider the contribution of the tensor mesons $\Phi_{\mu\nu}$ to the axial currents correlator. This contribution may be represented as a Feynman diagram shown in the Fig.\ref{Fig:aTa}.
Straightforward evaluation of the tensor-axial coupling is
\begin{eqnarray}
&&2 i\epsilon_{\mu\nu\rho\lambda}\Phi_{\mu\nu}q_\rho a_\lambda(q)\times c_A,\\
&&c_A=-8 N_c\intp \frac{2\mu^3(p)+p M f(p) f'(p)(p^2-3\mu^2(p))}{(p^2+\mu^2(p))^2}\nonumber
\end{eqnarray}
and the total contribution of the diagram in the Fig.~\ref{Fig:aTa} is 
\begin{eqnarray}
&&Fig.~\ref{Fig:aTa}\sim 8 c_A^2 q^2 \left(g_{\mu\nu}-\frac{q_\mu q_\nu}{q^2}\right)\times\\ \nonumber
&&\left(1-\left(\frac{\beta}{\alpha \ave{\sigma}}\right)^2 2 N_c\intp\frac{M^2 F^4(p)\mu^2(p)}{(p^2+\mu^2(p))^2} \right)^{-1}
\end{eqnarray}
Thus we can see that this diagram  is just $O(q^2)$-correction to the axial correlator. 
Since we are interested only in the LO over $q^2$ (evaluation of $F_\pi$), we should not evaluate this diagram.
 
 \section{Gasser-Leutwyler couplings}
 \label{SectGassCoupl}
 According to~\cite{Gasser:1983yg}, the low-energy constants $\bar l_i$ of the chiral lagrangian may be extracted from the ${\cal O}(m)$-corrections to physical quantities, e.g.
 \begin{eqnarray}
 &&M_\pi^2=m_\pi^2\left(1-\frac{m_\pi^2}{32\pi^2 F^2}\bar l_3+{\cal O}(m_\pi^4)\right)\\
 &&F_\pi^2=F^2\left(1+\frac{m_\pi^2}{8\pi^2 F^2}\bar l_4+{\cal O}(m_\pi^4)\right),
 \end{eqnarray}
 where $M_\pi, F_\pi$ are the pion mass and decay constants, $m_\pi^2=2\,m\,B$ and $B, F$ are the phenomenological parameters of the chiral lagrangian. Using our results (\ref{Res:Fpi}, \ref{Res:Mpi}), we can obtain ($m$ is given in $GeV$)
\begin{widetext}
\begin{eqnarray}
&&\label{F2}F^2=0.00284777 N_c-0.000868917+{\cal O}\left(\frac{1}{N_c}\right)\\
&&\label{B:value}B=1.7467+\frac{0.8183}{N_c}+{\cal O}\left(\frac{1}{N_c^2}\right)\\
\label{L_3}
 &&\bar l_3=\frac{-1.14251\,N_c\,\left( 1 + \frac{0.872354}{N_c} + \frac{0.874555\,\ln m}{N_c} +{\cal O}\left(\frac{1}{N_c^2}\right)\right) }{1 + \frac{0.936972}{N_c} +{\cal O}\left(\frac{1}{N_c^2}\right)}=0.0738267 - 1.14251\,N_c - 0.999\,\ln m+{\cal O}\left(\frac{1}{N_c}\right)\\
\label{L_4}
&&\bar l_4=\frac{-0.0793814\,N_c\,\left( 1 + \frac{0.232149}{N_c} + \frac{12.5977\,\ln m}{N_c} \right)}{1 + \frac{0.468486}{N_c}+{\cal O}\left(\frac{1}{N_c^2}\right)}=- 0.0793814\,N_c+0.0187608 - 1.000 \,\ln m+{\cal O}\left(\frac{1}{N_c}\right)
 \end{eqnarray}
\end{widetext}
 which gives
 \begin{eqnarray}
 F=88\, MeV,\;B=2.019\, GeV,\;\bar l_3=1.84,\;\bar l_4=4.98
 \label{l3l4:values}
 \end{eqnarray}
 at $m=0.0055\,\,GeV$, corresponding $M_\pi=0.142\,\,GeV$, $F_\pi=0.0937\,\,GeV$.

The values of $F, -\ave{\bar qq(m=0)}=-F^2 B$ in (\ref{l3l4:values}) were taken as input when we fixed the parameters $\rho, R$ in (\ref{Param:fit}).
Our values of ($\bar l_3, \bar l_4$) should be compared with the phenomenological estimates~\cite{Leutwyler:2006qq,Leutwyler:2007ae} as well as lattice predictions~\cite{Aubin:2004fs,DelDebbio:2006cn} given in Table~\ref{Table:l_i}.

\begin{table}[h]
\begin{tabular}{|c|c|c|c|c|c|}
\hline
{}&{$\chi PT$}&{MILC}&{Del Debbio}&{ETM}&{Our}\\
&\cite{Colangelo:2001df,Gasser:1983yg,Leutwyler:2007ae}&\cite{Bernard:2006zp}&{\textit{et. al.}~\cite{DelDebbio:2006cn}}&\cite{Boucaud:2007uk}&{prediction}\\
\hline
{$\bar l_3$}&{$2.9\pm 2.4$}&{$0.6\pm1.2$}&{$3.0\pm0.5$}&{$3.62\pm0.12$}&{$1.84$}\\
{$\bar l_4$}&{$4.4\pm 0.2$}&{$3.9\pm0.5$}&{---}&{$4.52\pm0.06$}&{$4.98$}\\
\hline
\end{tabular}
\caption{\label{Table:l_i}Estimates and predictions of the low-energy constants. The first column contains phenomenological estimates, the next three columns are lattice results from different collaborations, the last column contains our results. The first four columns of the table are taken from~\cite{Leutwyler:2007ae}.}
\end{table}

Notice that in (\ref{L_3} - \ref{l3l4:values}) we keep ${\cal O}(N_c, N_c^0)$-terms and drop NNLO terms ${\cal O}(1/N_c)$ in agreement with our general framework. Without such expansion one would get
\begin{eqnarray}
\bar l_3=0.28&&\bar l_4=4.28
\end{eqnarray}

\section{Chiral logs}
\label{SectChiralLogs}

Now we would like to discuss the chiral logarithms $m\ln m$  generated by pion loops in the effective action $\Gamma_{eff}$. First of all, we should notice that we are making not just a chiral expansion, but a double expansion over the $1/N_c$  in large-$N_c$ limit and the chiral expansion over the quark mass $m$. We must take this fact into account when checking the theorems-we should keep only LO and the NLO corrections. Since chiral logs originate from pion loops, which are ${\cal O}(1/N_c)$-corrections, the general chiral log theorem in our approximation has a form
\begin{eqnarray}
G(M_\pi)=G(M_\pi=0)\left(1+\gamma \frac{M_{\pi, LO}^2 \ln M_{\pi, LO}^2}{F_{\pi, LO}^2}+...  \right)\nonumber
\end{eqnarray}
where $\gamma$ is some specific numerical coefficient, and we took into account that $F_{\pi, LO}^2\sim N_c$.

Chiral logs have two sources:
\begin{enumerate}
\item Shift of dynamical quark mass $\delta M$ due to meson loops ("indirect" meson loops contribution). In this region contribution to chiral logs is obtained acting with operator
$
\frac{\delta M}{M}\frac{\partial}{\partial M}
$
on the analytic LO expressions for the correlators.
\item Pion small-momentum $q\approx 0$ region in pion loops ("direct" contribution). In this region pion propagator may be approximated as \begin{eqnarray}
\Pi^{-1}_\pi(q)\approx F_{\pi,LO}^2 (M_{\pi,LO}^2+ q^2),\label{Pi:low-q2-expansion}
\end{eqnarray} vertices $A_{1,ij}$ and $A_{2,ij}$ --- as constants (independent of $q$), so 
\begin{eqnarray}
\intq \Pi(q) A_{\pi}(q)\approx A_{\pi}(q=0)\frac{M_{\pi, LO}^2 \ln M_{\pi, LO}^2}{16\pi^2}+...
\end{eqnarray}
\end{enumerate}
So evaluation of the chiral logs is straightforward but tedious.
 \par The first relation for the constituent quark mass $M$~\cite{Nikolov:1996jj}
 \begin{eqnarray}
 \label{ChiLog:M}
 M(m)=M_0\left(1-\frac{3 M_{\pi, LO}^2}{32 \pi^2 F_{\pi, LO}^2} \ln M_{\pi, LO}^2+{\cal O}(M_\pi^4)\right)
 \end{eqnarray}
 may be most easily obtained expanding the vacuum equation (\ref{gap1}) and collecting $1/N_c$-terms. 
 \begin{eqnarray}
 &&-\half\frac{M_1}{M_0}\left(M_0\frac{\partial}{\partial M_0}\right)\Tr[Q(p)]+\\
 &&\nonumber\frac{1}{2}\intq \sum_i\left(V_2^i(q)-V_3^i(q)\right)\tilde\Pi_i(q)=0
 \end{eqnarray}
 Using (\ref{Pi:low-q2-expansion}) and the relations
 \begin{eqnarray}
 &&\intq \frac{1}{q^2+M_{\pi, LO}^2}=...+\frac{M_{\pi, LO}^2}{16\pi^2}\ln M_{\pi, LO}^2+{\cal O}(M_{\pi}^4)\nonumber\\
 &&\left(M_0\frac{\partial}{\partial M_0}\right)\Tr[Q(p)]=16 N_c \intp\frac{M^2 f^4(p)p^2}{(p^2+\mu^2(p))^2}\nonumber\\
 &&V^\phi_2(0)-V^\phi_2(0)=-8 N_c \intp\frac{M^2 f^4(p)p^2}{(p^2+\mu^2(p))^2},\nonumber
 \end{eqnarray}
 we can check that chiral log in $M_1$ satisfies (\ref{ChiLog:M}).
 \par In completely the same way, expanding Eqs.~(\ref{cond}) and using (\ref{ChiLog:M}), we can obtain for the chiral log in quark condensate
\begin{eqnarray}
\label{ChiLog:qq}
\langle\bar qq(m)\rangle=\langle\bar qq\rangle_0\left(1-\frac{3 M_{\pi, LO}^2}{32 \pi^2 F_{\pi, LO}^2} \ln M_{\pi, LO}^2+{\cal O}(M_{\pi}^4)\right)
 \end{eqnarray}
 
Similar evaluation of the chiral logs in $F_\pi$ and $M_\pi$ yields
 \begin{widetext}
 \begin{eqnarray}
 \label{ChiLog:Fpi}
 &&F_\pi^2(m)=F^2(0)\left(1-\frac{M_{\pi, LO}^2}{8\pi^2 F_{\pi, LO}^2} \ln M_{\pi, LO}^2+{\cal O}(M_{\pi}^4)\right)\\
 \label{ChiLog:Mpi}
 &&M_{\pi, LO}^2(m)=M_{\pi,LO}^2\left(1+\frac{M_{\pi, LO}^2}{32\pi^2 F_{\pi, LO}^2} \ln M_{\pi, LO}^2+{\cal O}(M_{\pi}^4)\right)
 \end{eqnarray}
 \end{widetext}
 in correspondence with low-energy theorems~\cite{LangackerPagels73} and large-$N_c$ expansion. From (\ref{ChiLog:Fpi},\ref{ChiLog:Mpi}) we can immediately see that the chiral logs in the low-energy couplings $\bar l_i$ are correct~\cite{Gasser:1983yg}.

 \section{Path Independence}
 \label{SectPathIndep}
 In our evaluations for the sake of convenience we chose the straight-line path in the transporter~(\ref{transporter}). Now we would like to demonstrate that the results of our evaluations are path independent, though on the intermediate steps we have explicit path dependence as a consequence of the zero-mode approximation~(\ref{Si}).\\
 Let us consider first the difference of the path integrals  over two contours: $\delta\left(\int v_\mu d\xi_\mu\right)=\oint v_\mu d\xi_\mu$. Due to Stock's theorem, this integral is reducible to the surface integral
 \begin{eqnarray}
 \label{PathIntegral}
 \oint v_\mu d\xi_\mu=\int dS_{\mu\nu}v_{\mu\nu}(x)
 \end{eqnarray}
 where for the monochromatic field $v_\mu(\xi)=v_\mu(q)e^{iq\cdot\xi}$ we have 
 \begin{eqnarray}
 \label{FieldStrength}
 &&v_{\mu\nu}(\xi)=v_{\mu\nu}(q)e^{iq\cdot \xi}\\
 &&v_{\mu\nu}(q)=i(q_\mu v_\nu(q)-q_\nu v_\mu(q)).
 \end{eqnarray}
  Since we are interested only in ${\cal O}(q^0)$-terms, the field strength is zero and thus integral over any closed contour is zero
 \begin{eqnarray}
 \label{Path_Integral}
 \oint v_\mu d\xi_\mu\approx v_{\mu\nu}(q)\int dS_{\mu\nu}=v_{\mu\nu}(q)S_{\mu\nu}=0
 \end{eqnarray} 
 For evaluation of $F_\pi$ from the $\langle j^{A,i}_\mu j^{A,j}_\nu\rangle$-correlator we can take $a_\mu=const$, which yields the path-independent result.
 \\
 Notice that the path independence exists only for small $q$, whereas for arbitrary $q$ (e.g. for the "dynamical" magnetic susceptibility considered in~\cite{Dorokhov:2005pg}) the path dependence exists and there is no other argument except simplicity why one should choose the straight-line path in the transporter~(\ref{transporter}).

\section{Discussion}
\label{SectDiscussion}
The aim of our work was the study of the pion physics beyond the chiral limit in the framework of the instanton vacuum model. We found the generating functional of the hadronic correlators with account of ${\cal O}(1/N_c,\,m,\,m/N_c,\,m \,\ln m/N_c)$-corrections and exploited it for evaluation of the corrections to different physical observables. The corrections considered in this paper include meson loops, finite width of instanton size distribution and quark-quark tensor interactions term.  In contrast to the expectations, we found that numerically the $1/N_c$-corrections to dynamical quark mass are large and mostly come from meson loops. As a consequence, we have large $1/N_c$-corrections to all the other quantities. To provide the values of $F_\pi(m=0), \ave{\bar qq(m=0)}$ in agreement with $\chi$PT, we offer a new set of parameters $\rho,\, R$~\ref{Param:fit}. Remarkably, this set of parameters is still in agreement with current phenomenological and lattice estimates~(\ref{rhoR},\ref{fromNegele0605256}).
\par The main result of this paper is the evaluation of the $F_\pi(m)$ and $M_\pi(m)$-dependence with account of $O(1/N_c, m, m/N_c, m/N_c \ln m)$-corrections. From comparison with $\chi PT$ we extract the values of the low energy constants $\bar l_3, \bar l_4$. Our results for the values of $\bar l_3$ and $\bar l_4$ are in a satisfactory agreement with phenomenological as well as lattice estimates (See Table~\ref{Table:l_i}). This means that the instanton vacuum is applicable for understanding of the low-energy physics, at least on the qualitative level. Evaluation of the other LEC's is in progress.
 \section*{Acknowledgements}
We would like to thank P. Pobylitsa for seminal discussions on different stages of the work. We also would like to thank P. Bowman for providing us with the lattice data on $M(m)$-dependence.
 The work has been partially supported by the DFG-Graduiertenkolleg Dortmund-Bochum, by the Verbundforschung of BMBF and by the bilateral Funds DFG-436 USB 113/11/0-1 between Germany and Uzbekistan.
\appendix 
 \section{Nonlocal vertices}
\label{SectVertices}
 \par One of the important features of the chiral quark model is the nonlocal interaction of the quarks with  mesons in (\ref{Z2}). On the one hand, due to the formfactors all the loop integrals in the model are convergent with an effective cut-off at the scale $p\sim\rho^{-1}$. On the other hand, such regularization explicitly violates gauge invariance. The modification of the quark-meson coupling to  
 \begin{eqnarray}
 \bar L(x, z)F(x-z)\Phi _i(z)\Gamma _i L(z, y)F(z-y)
 \label{InteractionVertex}
 \end{eqnarray}
 where $L$-factors are defined in (\ref{transporter}), formally restores the gauge covariance. However, it introduces unphysical path dependence of all correlators with vector and axial currents. In some of the simplest cases (e.g., constant field $a_\mu$) it is a trivial matter to show that the path dependence drops out, but it doesn't in general case. In our evaluations we choose conventional straight line path.
Let us consider for definiteness the fields of the plane wave form
\begin{eqnarray}
v_\mu(x)=v_\mu(q)e^{i q x}\\
a_\mu(x)=a_\mu(q)e^{i q x}
\end{eqnarray}
The integral along the straight-line path for such fields equals
\begin{eqnarray}
\int_z^x d\xi_\mu v_\mu(\xi)=(x-z)_\mu
v_\mu(q)e^{i q z}\frac{e^{iq\cdot(x-z)}-1}{iq\cdot(x-z)}
\end{eqnarray}
and the same for axial field $a$. 
In general the vertex~(\ref{InteractionVertex}) contains an infinite series of the terms $v,a$.
However in the next sections we'll need the vertices only up to $O[v^2, a^2]$-terms and restrict ourselves with this accuracy in this section.
Then the $L$-factors may be rewritten as
\begin{eqnarray}
&&L(z, y)=1+V_\mu(q)i(z-y)_\mu\frac{e^{iq\cdot(z-y)}-1}{iq\cdot(z-y)}+\nonumber\\
&&\nonumber V_\mu(q_1)V_\nu(q_2)i(z-y)_\mu i(z-y)_\nu\times\\
&&\frac{e^{iq_1\cdot(z-y)}-1}{iq_1\cdot(z-y)}\frac{e^{iq_2\cdot(z-y)}-1}{iq_2\cdot(z-y)}+O(V^3)
\end{eqnarray}
\begin{eqnarray}
&&\bar L(x, z)=\gamma_0 L(z, x)^\dagger\gamma_0=\nonumber\\
&&1+\bar V_\mu(q)i(x-z)_\mu\frac{e^{iq\cdot(x-z)}-1}{iq\cdot(x-z)}+\nonumber\\
&&\nonumber \bar V_\mu(q_1)\bar V_\nu(q_2)i(x-z)_\mu i(x-z)_\nu\times\\
&&\frac{e^{iq_1\cdot(x-z)}-1}{iq_1\cdot(x-z)}\frac{e^{iq_2\cdot(x-z)}-1}{iq_2\cdot(x-z)}+O(V^3),
\end{eqnarray}
where $V_\mu(q)=v_\mu(q)+a_\mu(q)\gf, \bar V_\mu(q)=v_\mu(q)-a_\mu(q)\gf$.
 The interaction vertices may be compactly rewritten in terms of the functions 
\begin{eqnarray}
&& G_\mu(x-z;q):=i(x-z)_\mu\frac{e^{iq\cdot(x-z)}-1}{iq\cdot(x-z)}F(x-z) \\
&& H_{\mu\nu}(x-z;q_1, q_2):=i(x-z)_\mu i(x-z)_\nu\times\\
&&\frac{e^{iq_1\cdot(x-z)}-1}{iq_1\cdot(x-z)}\frac{e^{iq_2\cdot(x-z)}-1}{iq_2\cdot(x-z)}F(x-z)\nonumber
\end{eqnarray}
In $p$-space Fourier expansions of these functions are
\begin{eqnarray}
&& G_\mu(p;q):=\int d^4x\, G_\mu(x;q)e^{ip\cdot x}=\\
&&\sum^\infty_{n=0} \frac{1}{(n+1)!}F_{\mu, \mu_1\ldots \mu_n}(p)q_{\mu_1}\ldots  q_{\mu_n}\nonumber \\
&& H_{\mu\nu}(p;q_1, q_2):=\int d^4x\, H_{\mu\nu}(x;q_1, q_2)e^{ip\cdot x}=\\ \nonumber
&&\sum^\infty_{n, k=0} \frac{1}{(n+1)!(k+1)!}F_{\mu, \nu, \mu_1\ldots \mu_n, \nu_1\ldots \nu_k}(p)\times\\
&&q_{1_{\mu_1}}\ldots q_{1_{\mu_n}}q_{2_{\nu_1}}\ldots  q_{2_{\nu_k}}\nonumber
\end{eqnarray}

\begin{figure}[h]
  \centering
\includegraphics[scale=0.25]{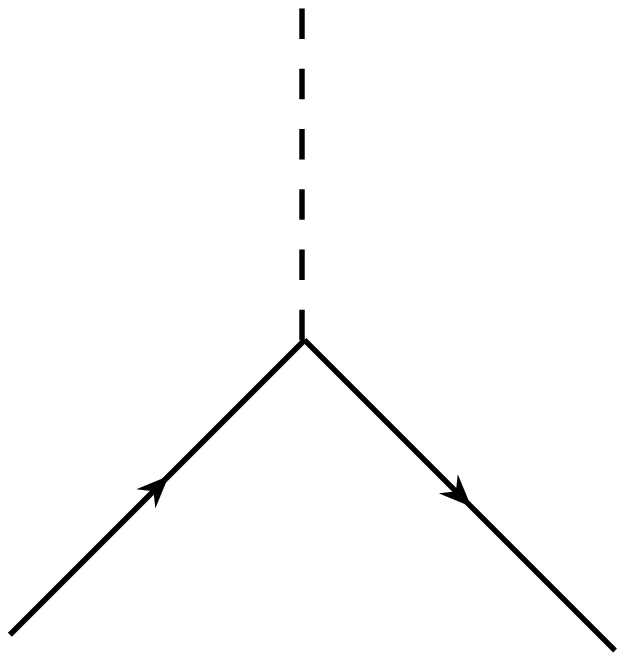}
\includegraphics[scale=0.25]{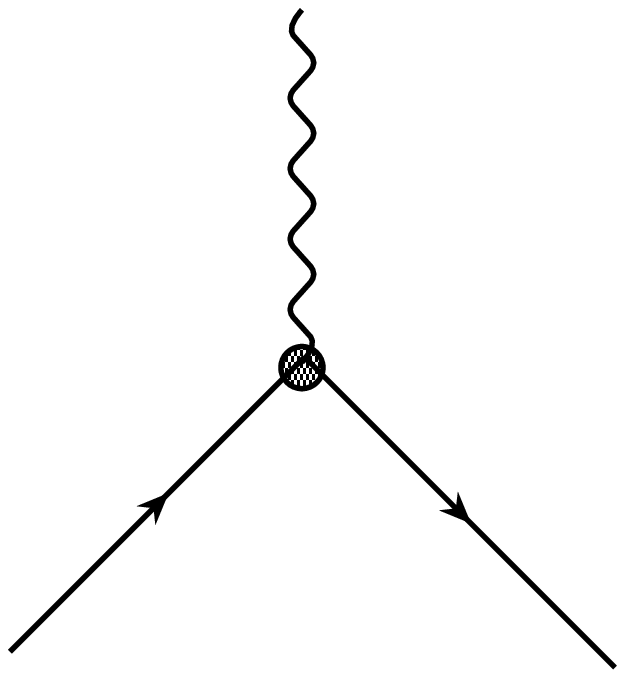}
\includegraphics[scale=0.25]{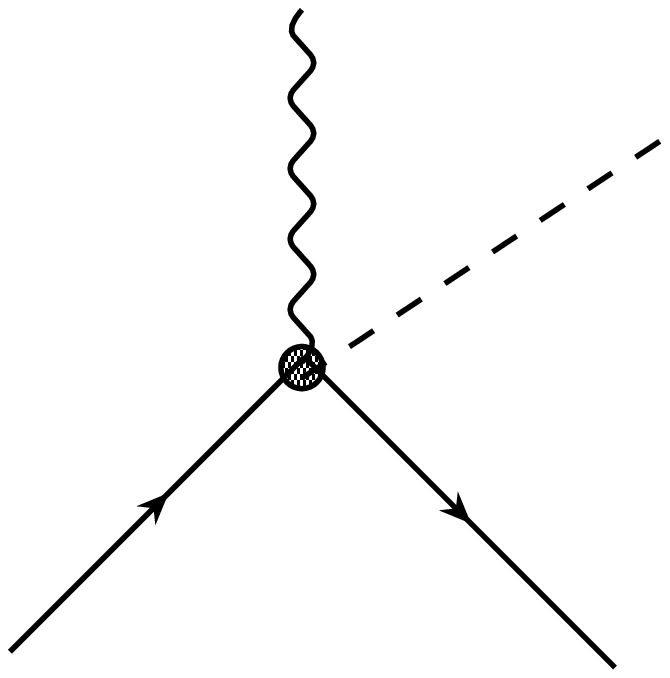}
\includegraphics[scale=0.25]{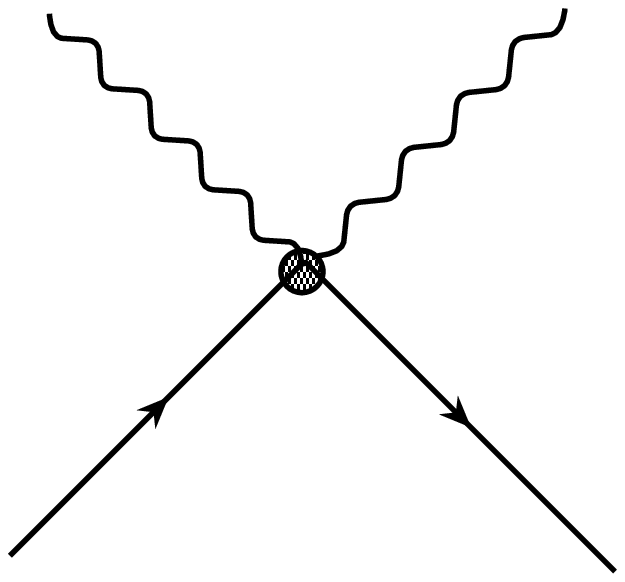}
\includegraphics[scale=0.25]{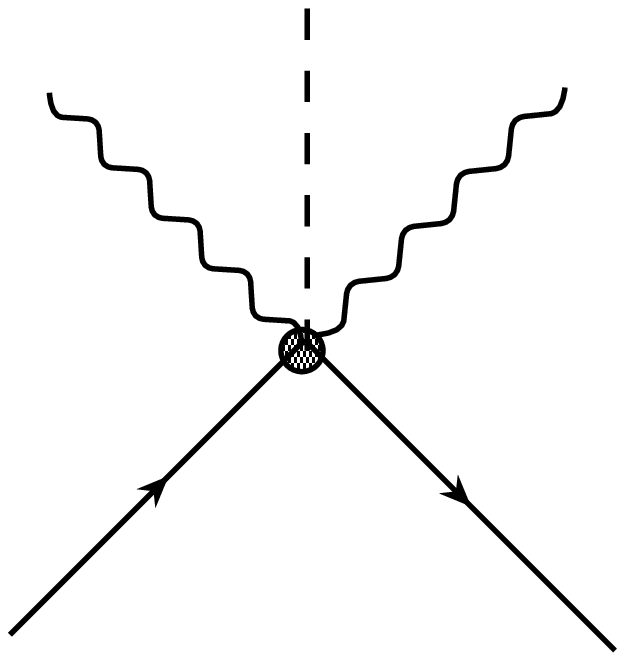}
\caption{The basic vertices (Up to ${\cal O}(v^2, a^2)$-terms). The wavy lines correspond to the currents $v_\mu, a_\mu$, the dashed line corresponds to the meson, the bulbs correspond to all the possible (local and nonlocal) couplings of the current $V$ to the constituent quarks.}\label{fig:Vertices}
\end{figure}

In terms of these functions interaction terms may be rewritten as
($\Phi_i(x)=\Phi_i(k)e^{ikx}$, integration over the dummy momenta $k, q, q_1, q_2$ is implied)
\begin{widetext}
\begin{eqnarray}
&&\nonumber\bar L(x, z)F(x-z)\Phi _i(z)\Gamma _i L(z, y)F(z-y)=\\
&&[G_\mu(p, q)\Gamma_i F(p+k+q)+F(p)\Gamma_i G_\mu (p+k+q, -q)]v_\mu(q)\Phi_i(k) +\\
&&\nonumber\left[\frac{}{}G_\mu(p, q_1)\Gamma_i G_\nu (p+k+q_1+q_2, -q_2)+\right. \\
&&\nonumber\left. \half\left(H_{\mu\nu}(p, q_1, q_2)\Gamma_iF(p+k+q_1+q_2)+
F(p)\Gamma_iH_{\mu\nu}(p+k+q_1+q_2;-q_1, -q_2)\right)\right]v_\mu(q_1)v_\nu(q_2)\Phi_i(k)+\\
&&[-G_\mu(p, q)\Gamma_i F(p+k+q)+F(p)\Gamma_i G_\mu (p+k+q,-q)]a_\mu(q)\gf\Phi_i(k) +\\
&&\nonumber\left[\frac{}{}-G_\mu(p, q_1)\Gamma_i G_\nu (p+k+q_1+q_2, -q_2)+\right. \\
&&\nonumber\left. \half\left(H_{\mu\nu}(p, q_1, q_2)\Gamma_iF(p+k+q_1+q_2)+
F(p)\Gamma_iH_{\mu\nu}(p+k+q_1+q_2;-q_1,-q_2)\right)\right]a_\mu(q_1)a_\nu(q_2)\Phi_i(k)\\
&&+ O(av)-mixing\, \, terms\, \, we\, \, are\, \, not\, \, interested\, \, in\, \, now
\end{eqnarray}
\par Due to chiral symmetry breaking field $\sigma$ has nonzero vacuum expectation value (VEV). 
This generates momentum dependent constituent quark mass $\mu(p)$ and also generates new nonlocal vertices
we should take into account in expansions over the fields $\Phi$ and currents $v, a$:
\begin{eqnarray}
&&V_\sigma= \sigma_{vac}\lbrace
[G_\mu(p, q) F(p+q)+F(p) G_\mu (p+q,-q)]v_\mu(q) +\\
&&\nonumber\left[\frac{}{}G_\mu(p, q_1) G_\nu (p+q_1+q_2,-q_2)+\right. \\
&&\nonumber\left. \half\left(H_{\mu\nu}(p, q_1, q_2)F(p+q_1+q_2)+
F(p)H_{\mu\nu}(p+q_1+q_2;-q_1, -q_2)\right)\right]v_\mu(q_1)v_\nu(q_2)+\\
&&\label{vertices}[-G_\mu(p, q) F(p+q)+F(p) G_\mu (p+q, -q)]a_\mu(q)\gf +\\
&&\nonumber\left[\frac{}{}-G_\mu(p, q_1) G_\nu (p+q_1+q_2, -q_2)+\right. \\
&&\nonumber\left. \half\left(H_{\mu\nu}(p, q_1, q_2)F(p+q_1+q_2)+
F(p)H_{\mu\nu}(p+q_1+q_2;-q_1, -q_2)\right)\right]a_\mu(q_1)a_\nu(q_2)\rbrace
\end{eqnarray}
\end{widetext}
The vertices which are generated are shown on the Fig.\ref{fig:Vertices}
 \section{Finite width corrections}
 \label{SectFWCDetails}
 In this section we give solution of the vacuum equations (\ref{Gap:FWC}) in the small-width approximation.
  In this approximation we can make a systematic expansion to find the unknown function $\Phi(\rho)$ and parameter $\lambda$. We do not specify the shape of $d(\rho)$, since all the finite width corrections will depend only on the width $\delta\rho^2$ defined in (\ref{SizeDefinition}). 
 Expansion over $\delta\rho^2$ is straightforward :
 \begin{widetext}
 \begin{eqnarray}
 &&\Phi_0(\rho):=\phi_0+\phi_1(\rho-\bar \rho)+\phi_2\frac{(\rho-\bar \rho)^2}{2}+...\\
 &&\tilde K(p,\rho):=\tilde K_0(p)+\tilde K_1(p)(\rho-\bar \rho)+\tilde K_2(p)\frac{(\rho-\bar \rho)^2}{2}+...\\
 &&\int d\rho\,d(\rho)=1\\
 &&\int \rho d\rho\,d(\rho)=\bar\rho\\
 &&\int (\rho-\bar\rho)^2 d\rho\,d(\rho)=\delta^2\\
 &&\Rightarrow 2\,\left(\phi_0^2+\delta^2(\phi_1^2+\phi_0\phi_2)\right)=\frac{N}{V},\label{NormalizationExpanded}\\
 &&\int d\rho d(\rho) \tilde K(p,\rho)\Phi_0(\rho)=\phi_0 \tilde K_0(p)+\frac{\delta^2}{2}(2\phi_1 \tilde K_1(p)+\phi_0 \tilde K_2(p)+\phi_2 \tilde K_0(p))\\
 &&\Rightarrow \phi_0 =\frac{1}{4} Tr_{p,\{f,c\}}\left(\frac{i c\, \lambda^{0.5} \tilde K_0(p)}{\hat p+i m +i c\, \lambda^{0.5} (\tilde K_0(p)\phi_0+\frac{\delta^2}{2}(2\phi_1 \tilde K_1(p)+\phi_0 \tilde K_2(p)+\phi_2 \tilde K_0(p)))}\right)\label{phi0}\\
 &&\Rightarrow \phi_1 =\frac{1}{4} Tr_{p,\{f,c\}}\left(\frac{i c\, \lambda^{0.5} \tilde K_1(p)}{\hat p+i m +i c\, \lambda^{0.5} (\tilde K_0(p)\phi_0+\frac{\delta^2}{2}(2\phi_1 \tilde K_1(p)+\phi_0 \tilde K_2(p)+\phi_2 \tilde K_0(p)))}\right)\label{phi1}\\
 &&\Rightarrow \phi_2 =\frac{1}{4} Tr_{p,\{f,c\}}\left(\frac{i c\, \lambda^{0.5} \tilde K_2(p)}{\hat p+i m +i c\, \lambda^{0.5} (\tilde K_0(p)\phi_0+\frac{\delta^2}{2}(2\phi_1 \tilde K_1(p)+\phi_0 \tilde K_2(p)+\phi_2 \tilde K_0(p)))}\right)\label{phi2}
 \end{eqnarray}
 Equations (\ref{NormalizationExpanded},\ref{phi0}-\ref{phi2}) form a system of four equations with four independent variables $\phi_0-\phi_2,\lambda$. By definition, $$\phi_n=\frac{\Phi^{(n)}_0(\bar \rho)}{n!}.$$ Note that \textit{none of these variables is suppressed over $\delta$}. We can make the next step and expand over the variable $\delta$:
 \begin{eqnarray}
 &&\phi_m:=\sum_n \phi_{mn}\frac{\delta^n}{n!}\\
 &&\lambda:=\sum_n \lambda_{n}\frac{\delta^n}{n!}\\
 &&\lambda^{0.5}=\lambda_{0}^{0.5}\left(1+\delta\half\frac{\lambda_1}{\lambda_0}+\delta^2\left(\half\frac{\lambda_2}{\lambda_0}-\frac{\lambda_1^2}{8\,\lambda_0^2}\right)+...\right):=\sum_n l_n \frac{\delta^n}{n!}
 \end{eqnarray}
 From the normalization condition we get
 \begin{eqnarray}
 &&2\phi_{00}^2=\frac{N}{V}\mbox{  \textit{In agreement with $\delta=0$ case}}\label{NormalizationExpanded1}\\
 &&\phi_{01}=0\label{NormalizationExpanded2}\\
 &&\phi_{00}\phi_{02}+\phi_{10}^2+\phi_{00}\phi_{20}=0\label{NormalizationExpanded3}
 \end{eqnarray}
 Equations (\ref{phi0}-\ref{phi2}) give
 \begin{eqnarray}
 &&\Rightarrow \phi_{00} =\frac{1}{4} Tr_{p,\{f,c\}}\left(\frac{i c\, l_0 \tilde K_0(p)}{\hat p+i m +i c\, l_0 \tilde K_0(p)\phi_{00}}\right)\mbox{  \textit{In agreement with $\delta=0$ case}}\label{phi_00}\\
 &&0=\phi_{01} =\frac{1}{4} Tr_{p,\{f,c\}}\left(\frac{i c\, l_1 \tilde K_0(p)}{\hat p+i m +i c\, l_0 \tilde K_0(p)\phi_0}\right)+\frac{1}{4} Tr_{p,\{f,c\}}\left(\frac{ c^2\, l_0 \tilde K_0^2(p)(\phi_{00}l_1+\phi_{01}l_0)}{(\hat p+i m +i c\, l_0 \tilde K_0(p)\phi_{00})^2}\right)=\nonumber\\
 &&l_1\left(\frac{1}{4} Tr_{p,\{f,c\}}\left(\frac{i c\,  \tilde K_0(p)}{\hat p+i m +i c\, l_0 \tilde K_0(p)\phi_0}\right)+\frac{1}{4} Tr_{p,\{f,c\}}\left(\frac{ c^2\, l_0 \tilde K_0^2(p)\phi_{00}}{(\hat p+i m +i c\, l_0 \tilde K_0(p)\phi_{00})^2}\right)\right)\label{phi_01}\\
 &&\mbox{$\Rightarrow l_1=0$ since expression inside the brackets is not zero $\Rightarrow \lambda_1=0$}\nonumber\\
 &&\Rightarrow \phi_{02} =\frac{1}{4} Tr_{p,\{f,c\}}\left(\frac{i c\, l_2 \tilde K_0(p)}{\hat p+i m +i c\, l_0 \tilde K_0(p)\phi_{00}}\right)+\label{phi_02}\\
 &&\frac{1}{8} Tr_{p,\{f,c\}}\left(\frac{ c^2\,l_0\tilde K_0(p) (l_2 \phi_{00}\tilde  K_0(p)+l_0(\phi_{02}\tilde  K_0(p)+2\phi_{10}\tilde  K_1(p)+\tilde  K_2(p)\phi_{00}+\phi_{20}\tilde  K_0(p))}{(\hat p+i m +i c\, l_0 \tilde K_0(p)\phi_{00})^2}\right)\nonumber\\
 &&\Rightarrow \phi_{10} =\frac{1}{4} Tr_{p,\{f,c\}}\left(\frac{i c\, l_0 \tilde K_1(p)}{\hat p+i m +i c\, l_0 \tilde K_0(p)\phi_{00}}\right)\label{phi_10}\\
 &&\phi_{11} =\frac{1}{4} Tr_{p,\{f,c\}}\left(\frac{i c\, l_1 \tilde K_1(p)}{\hat p+i m +i c\, l_0 \tilde K_0(p)\phi_0}\right)+\frac{1}{4} Tr_{p,\{f,c\}}\left(\frac{ c^2\, l_0 \tilde K_0(p)\tilde K_1(p)(\phi_{00}l_1+\phi_{01}l_0)}{(\hat p+i m +i c\, l_0 \tilde K_0(p)\phi_{00})^2}\right)=0\label{phi_11}\\
 &&\mbox{since $\lambda_1=0,\phi_{01}=0$}\nonumber\\
 &&\Rightarrow \phi_{12} =\frac{1}{4} Tr_{p,\{f,c\}}\left(\frac{i c\, l_2 \tilde K_1(p)}{\hat p+i m +i c\, l_0 \tilde K_0(p)\phi_{00}}\right)+\label{phi_12}\\
 &&\frac{1}{8} Tr_{p,\{f,c\}}\left(\frac{ c^2\,l_0\tilde K_1(p) (l_2 \phi_{00}\tilde  K_0(p)+l_0(\phi_{02}\tilde  K_0(p)+2\phi_{10}\tilde  K_1(p)+\tilde  K_2(p)\phi_{00}+\phi_{20}\tilde  K_0(p))}{(\hat p+i m +i c\, l_0 \tilde K_0(p)\phi_{00})^2}\right)\nonumber\\
 &&\Rightarrow \phi_{20} =\frac{1}{4} Tr_{p,\{f,c\}}\left(\frac{i c\, l_0 \tilde K_2(p)}{\hat p+i m +i c\, l_0 \tilde K_0(p)\phi_{00}}\right)\label{phi_20}\\
 &&\phi_{21}=\frac{1}{4} Tr_{p,\{f,c\}}\left(\frac{i c\, l_1 \tilde K_2(p)}{\hat p+i m +i c\, l_0 \tilde K_0(p)\phi_0}\right)+\frac{1}{4} Tr_{p,\{f,c\}}\left(\frac{ c^2\, l_0 \tilde K_0(p)\tilde K_2(p)(\phi_{00}l_1+\phi_{01}l_0)}{(\hat p+i m +i c\, l_0 \tilde K_0(p)\phi_{00})^2}\right)=0\label{phi_21}\\
 &&\mbox{since $\lambda_1=0,\phi_{01}=0$}\nonumber\\
 &&\Rightarrow \phi_{22} =\frac{1}{4} Tr_{p,\{f,c\}}\left(\frac{i c\, l_2 \tilde K_2(p)}{\hat p+i m +i c\, l_0 \tilde K_0(p)\phi_{00}}\right)+\label{phi_22}\\
 &&\frac{1}{8} Tr_{p,\{f,c\}}\left(\frac{ c^2\,l_0\tilde K_2(p) (l_2 \phi_{00}\tilde  K_0(p)+l_0(\phi_{02}\tilde  K_0(p)+2\phi_{10}\tilde  K_1(p)+\tilde  K_2(p)\phi_{00}+\phi_{20}\tilde  K_0(p)))}{(\hat p+i m +i c\, l_0 \tilde K_0(p)\phi_{00})^2}\right)\nonumber
 \end{eqnarray}
 Thus we finish with a definite set of equations sufficient to determine all the constants $\phi_{mn}, l_n$.
 \begin{enumerate}
 \item Evaluate the "ordinary" LO vacuum equation (\ref{NormalizationExpanded1},\ref{phi_00}). Obtain the values $l_0,\, \phi_{00}$
 \item Substitute these values into (\ref{phi_10},\ref{phi_20}) and evaluate $\phi_{10},\phi_{20}$
 \item From (\ref{NormalizationExpanded3}) evaluate $\phi_{02}$ 
 \item Consider (\ref{phi_02}) as a (linear !) equation w.r.t. $l_2$ and get the value $l_2$
 \item Evaluate directly the remaining values $\phi_{12},\phi_{22}$
 \end{enumerate}
 Notice that so far we haven't used any particular form of the size distribution $\rho$. The only assumption we did is that the width is small, $\delta\rho^2 \ll \bar\rho^2$
 \end{widetext}
 \section{Structure of the $\ave{j^{5,a}_\mu j^{5,b}_\nu}$-correlator}
 \label{Sect:Transversity}
 In this section we are going to prove that due to the chiral symmetry correlator $\ave{j^{5,a}_\mu j^{5,b}_\nu}$ has a form
\begin{eqnarray}
\ave{j^{5,a}_\mu j^{5,b}_\nu}=F_\pi^2 \delta^{ab}\left[\delta_{\mu\nu}-\frac{q_\mu q_\nu}{q^2+ M_\pi^2}\right]+{\cal O}(q^2)\label{Transversity:structure}
\end{eqnarray}

The easiest way is to start from (\ref{S}), split the field $\Phi$ as $\Phi=\sigma U+\Phi'$ and make the chiral rotation of the quark fields
\begin{eqnarray}
\psi(x)\to U^{\gamma_5,1/2,\dagger}(x)\psi(x),\;\psi^{\dagger}(x)\to U^{\gamma_5,1/2,\dagger}(x)\psi^\dagger(x)
\label{psi:Rotation}
\end{eqnarray}
where the matrix $U$ was defined in (\ref{U:definition}) and will be parameterized as
\begin{eqnarray}
U(x)=\exp \left(i \vec u \vec \tau\right)=1+i\vec u \vec \tau-\frac{\vec u^2}{2}+...
\end{eqnarray}
The transformation (\ref{psi:Rotation}) has Jacobian equal to one, so we have just to evaluate the rotation of the Dirac operator in $S$. This is convenient to introduce a special notation
\begin{eqnarray}
&&i^n x_{\mu_1}...x_{\mu_n}F(x)=F_{\mu_1,...,\mu_n}(x)=\\
&&\intp e^{-ip\cdot (x-z)}F_{, \mu_1,...,\mu_n}(p),
\end{eqnarray}
\textit{i.e.} \textit{for formfactors the lower index corresponds to differentiation in momentum space} (not in coordinate !) whereas for all the other quantities we will use the lower index for differentiation in coordinate space, \textit{e.g.}
\begin{eqnarray}
u_{,\mu}(x)\equiv \frac{\partial u(x)}{\partial x^\mu}=\frac{\partial u(x)}{\partial x_\mu}
\end{eqnarray}
One more agreement, for the sake of brevity sometimes we will not write out explicitly the flavour dependence for the quantities $\vec u$ and $\vec a_\mu$, implying
\begin{eqnarray}
u\equiv \frac{\vec u \vec \tau}{2},\; a_\mu\equiv \frac{\vec a_\mu \vec \tau}{2},
\end{eqnarray}
Now we are going to evaluate the result of rotation. Note that actually we are making a double expansion: up to the second order in $\vec u\sim \vec a_\mu$ and up to the second order in derivatives of the field $u$ (assuming $a_\mu$ to be the first order).
Rotation of the local part is quite trivial, the result is 
\begin{eqnarray}
&&U^{\gf, 1/2\dagger}\left(\hat p+\hat a \gf +im\right)U^{\gf, 1/2\dagger}=...=\hat p +\hat a \gf + u_{,\mu}\gamma^\mu\gf+\nonumber\\
&&\frac{\vec \tau}{2}\cdot (\vec a_\mu+ \vec u_{,\mu})\times \vec u+im\left(1-i u\gf-\frac{u^2}{2}\right)
\end{eqnarray}
Notice that the term containing cross-product (in isospace) doesn't contribute since it is already a highest order in $\vec a, \vec u$ which we keep and it should be combined with another vector-isovector combined of $\vec u, \vec a_\mu$ and the derivatives.
For the nonlocal part rotation is a bit more tricky since we have matrices $U$ at three different points:
\begin{widetext}
\begin{eqnarray}
\label{Dirac}
&&iM\bar L(x,z) F(x-z)U^{\gf}(z) L(z,y) F(z-y)+v\rightarrow 
iM U^{\gamma_5,1/2,\dagger}(x)\bar L(x,z) F(x-z)U^{\gamma_5}(z)L(z,y) F(z-y)U^{\gamma_5,1/2,\dagger}(y)\nonumber
+v',
\\\nonumber
&& v'=\frac{iM}{\sigma}
U^{\gamma_5,1/2,\dagger}\bar L F(p)U^{\gamma_5,1/2}\Gamma\cdot \Phi^{''}
U^{\gamma_5,1/2}LF(p)U^{\gamma_5,1/2,\dagger}.
\end{eqnarray}
\end{widetext}
where meson fluctuations $\Phi^{''}_i$ and $\Phi^{'}_i$ are related by the chiral rotation,
\begin{eqnarray}
\Phi^{''}=U^{\gamma_5,1/2,\dagger}\Phi'U^{\gamma_5,1/2,\dagger}.
\end{eqnarray}
for a moment we will consider only the LO result and ignore the fluctuations term $v$.

First, we are going to make the Taylor expansion at the point $z$,
\begin{widetext}
\begin{eqnarray}
&&U^{\gf,1/2,\dagger}(x)=\left(1-i u(z)\gf-\frac{\vec u^2(z)}{8}\right)+\left(-i u_{,\mu}(z)\gf -\frac{\vec u(z)\vec u_{,\mu}(z)}{4}\right)(x-z)_\mu+\\
&&\left(-\frac{i}{2} u_{\mu,\nu}(z)\gf+\frac{\vec u_{,\mu}(z)\vec u_\nu(z)}{4}+\frac{\vec u(z)\vec u_{\mu\nu}(z)}{4}\right)(x-z)_\mu(x-z)_\nu\nonumber+{\cal O}(a^2, u^2, au, \partial^3)\\
&&U^{\gf,1/2,\dagger}(y)=\left(1-i u(z)\gf-\frac{\vec u^2(z)}{8}\right)+\left(i u_{,\mu}(z)\gf +\frac{\vec u(z)\vec u_{,\mu}(z)}{4}\right)(z-y)_\mu+\\
&&\left(-\frac{i}{2}u_{\mu,\nu}(z)\gf+\frac{\vec u_{,\mu}(z)\vec u_\nu(z)}{4}+\frac{\vec u(z)\vec u_{\mu\nu}(z)}{4}\right)(z-y)_\mu(z-y)_\nu\nonumber+{\cal O}(a^2, u^2, au, \partial^3)
\end{eqnarray}
Using the general expressions for the nonlocal vertices from the previous Section~\ref{SectVertices}, one can show that
\begin{eqnarray}
&&\bar L F(x-z)=F(x-z)- a_\mu(q)F_{\mu}(x-z)\gf+\half a_\mu(q) a_\nu(q) F_{\mu\nu}(x-z)+{\cal O}(a^3)\\
&&L F(z-y)=F(z-y)+ a_\mu(q)F_{\mu}(z-y)\gf+\half a_\mu(q) a_\nu(q) F_{\mu\nu}(z-y)+{\cal O}(a^3)\\
&&\Rightarrow \label{barLF:rotated}
U^{\gf,1/2,\dagger}(x) \bar L F(x-z) U^{\gf,1/2}(z)=F(x-z)-\left(a_\mu(q)\gf+u_{,\mu}(z)\gf+\frac{\vec \tau}{4}\vec u(z)\times (\vec a_\mu(q)+\vec u_{,\mu}(z)) \right)F_\mu(x-z)+\nonumber\\
&& \frac{1}{8} (\vec a_\mu(q)+\vec u_{,\mu}(z))(\vec a_\nu(q)+\vec u_\nu(z)) F_{\mu\nu}(x-z)+\frac{i}{2}u_{\mu\nu}\gf F(x-z)+{\cal O}(a^3, u^3, a^2 u, a u^2, \partial_\mu (\vec u \partial_\mu \vec u) )
\end{eqnarray}
Notice that we drop the terms $\partial_\mu (\vec u \partial_\mu \vec u)$ which give just a total derivative
\begin{eqnarray}
&&\Rightarrow \label{LF:rotated}
U^{\gf,1/2,\dagger}(z) \bar L F(z-y) U^{\gf,1/2}(y)=F(z-y)+\left(a_\mu(q)\gf+u_{,\mu}(z)\gf-\frac{\vec \tau}{4}\vec u(z)\times (\vec a_\mu(q)+\vec u_{,\mu}(z)) \right)F_\mu(z-y)+\nonumber\\
&& \frac{1}{8} (\vec a_\mu(q)+\vec u_{,\mu}(z))(\vec a_\nu(q)+\vec u_\nu(z)) F_{\mu\nu}(z-y)+\frac{i}{2}u_{\mu\nu}\gf F(z-y)+{\cal O}(a^3, u^3, a^2 u, a u^2, \partial_\mu (\vec u \partial_\mu \vec u) )\\
&&\Rightarrow U^{\gamma_5,1/2,\dagger}(x)\bar L(x,z) F(x-z)U^{\gamma_5}(z)L(z,y) F(z-y)U^{\gamma_5,1/2,\dagger}(y)=...=\\
&&F(x-z)F(z-y)+\left(a_\mu(q)\gf+u_{,\mu}(z)\gf+\frac{\vec \tau}{4}\vec u(z) \times (\vec a_\mu(q)+\vec u_{,\mu}(z)) \right) (F_\mu(x-z)F(z-y)-F(x-z)F_\mu(z-y))+\nonumber\\
&&\frac{1}{8} (\vec a_\mu (q) + \vec u_{,\mu}(z))(\vec a_\nu (q) + \vec u_\nu(z))(F_{\mu\nu}(x-z)F(z-y)+F(x-z)F_{\mu\nu}(z-y)-2 F_\mu(x-z)F_\nu(z-y))+\nonumber\\
&&\frac{i}{2}u_{\mu\nu}(z)\gf (F_{\mu\nu}(x-z)F(z-y)+F(x-z)F_{\mu\nu}(z-y))+{\cal O}(a^3, u^3,...)\nonumber
\end{eqnarray}
\end{widetext}
Note that
\begin{enumerate}
\item The term $\sim \vec a_\mu\times \vec u$ doesn't contribute due to its isovector structure
\item The term $\sim u_{\mu\nu}$ may only contribute beyond the chiral limit: it contains the second derivative, and the ``free'' $\vec u$ (without derivatives) comes only via ${\cal O}(m u)$-term.
\item All the remaining structures have a form $(\vec a_\mu(q)+\vec u_{,\mu}(z))$
\item Beyond the chiral limit expansion of $Tr\ln$ yields
\begin{eqnarray}
&&\Gamma_{eff}=\alpha_0 (\vec a_\mu+ \vec u_{,\mu})^2+m\,\alpha_1 \vec u_{,\mu} (\vec a_\mu+ \vec u_{,\mu})+m\,\alpha_2 \vec u^2=\nonumber\\
&&=\frac{1}{2}\left[F_{aa}^2 \vec a_\mu^2+F_{uu}^2 \vec u_{,\mu}^2+2 F_{au}^2 \vec a_\mu\vec u_{,\mu} + F_{uu}^2  M_\pi^2 \vec u^2\right] +\nonumber\\
&&+{\cal O}(a^3,u^3, m^2),
\end{eqnarray}
where the constants $F_{ij}$ differ only beyond chiral limit:
\begin{eqnarray}
F_{aa}^2 -F_{uu}^2=2\left(F_{au}^2 -F_{uu}^2\right)=-\alpha_1 \,m
\end{eqnarray}
\end{enumerate}
For evaluation of the meson loop contribution (term $v'$) one should notice that only the second order, $v^{'2}$, is essential. Using expansions (\ref{barLF:rotated},\ref{LF:rotated}), one can show that in the chiral limit the result depends only upon the structure $\vec a_\mu+\vec u_{,\mu}$. Beyond the chiral limit ${\cal O}(m)$-correction has a form $\sim m\, \vec u_{,\mu} (\vec a_\mu+ \vec u_{,\mu})$ This completes the proof that the correlator of two axial currents has a form~(\ref{Transversity:structure}).

  \end{document}